\documentclass[]{aastex}          
\usepackage{spr-astr-addons}    

\usepackage{color}

\usepackage{here}

\begin{document}
%
\title{The interstellar medium in young supernova remnants: key to the production of cosmic X-rays and $\gamma$-rays}

\shorttitle{The ISM in young SNRs}
\shortauthors{Sano \& Fukui}

\author{Hidetoshi Sano\altaffilmark{1}} 
\and 
\author{Yasuo Fukui\altaffilmark{2}} 

\altaffiltext{1}{National Astronomical Observatory of Japan, Mitaka, Tokyo 181-8588, Japan; hidetoshi.sano@nao.ac.jp}
\altaffiltext{2}{Department of Physics, Nagoya University, Furo-cho, Chikusa-ku, Nagoya 464-8601, Japan; fukui@a.phys.nagoya-u.ac.jp}

\begin{abstract}
We review recent progress in elucidating the relationship between high-energy radiation and the interstellar medium (ISM) in young supernova remnants (SNRs) {with ages of $\sim$2000} yr{, focusing in particular} on RX~J1713.7$-$3946 and RCW~86. {Both} SNRs {emit} strong nonthermal X-rays and TeV $\gamma$-rays, and {they contain} clumpy {distributions} of interstellar gas {that} includes {both} atomic and molecular hydrogen. We find that shock--cloud {interactions provide} a viable explanation {for} the spatial correlation between the X-rays and ISM. In {these interactions,} the supernova shocks hit the {typically pc-scale} dense cores{, generating a} highly turbulent velocity field {that} amplifies the magnetic field up to 0.1--1 mG. This amplification leads to enhanced nonthermal synchrotron emission around the clumps, whereas the cosmic-ray electrons do not penetrate the clumps. {Accordingly}, the nonthermal X-rays {exhibit a} spatial distribution similar to {that of} the ISM {on the} pc scale, while they are anticorrelated at sub-pc {scales}. {These results predict} that hadronic $\gamma$-rays can be emitted from the dense cores, resulting in {a} spatial correspondence {between} the $\gamma$-rays {and} the ISM. The current pc-scale resolution of $\gamma$-ray observations is too low to resolve {this} correspondence. Future $\gamma$-ray observations with {the} Cherenkov Telescope Array will {be able to} resolve the sub-pc-scale $\gamma$-ray distribution and provide {clues} to the {origin of these cosmic $\gamma$-rays}.
\end{abstract}


\section{Introduction}\label{s:introduction}
Cosmic-ray acceleration in supernova remnants (SNRs) is the most promising mechanism for {accelerating Galactic} cosmic rays, {which} mainly {comprise} relativistic protons {with energies less than} 10$^{15.5}$ eV. Considerable efforts have been {devoted to} theoretical works {to elucidate} the details of {particle acceleration} \citep[e.g.,][]{1978MNRAS.182..147B,1978ApJ...221L..29B}. Recent progress in X-ray and $\gamma$-ray observations has allowed us to explore the origin of cosmic rays. {Young} SNRs{, which ages} around {2000} yr{,} are of particular interest because they emit {higher-}energy X-rays and $\gamma$-rays {than} SNRs younger than 1000 yr \citep[e.g., G1.9$+$0.3 at $156\pm11$ yr,][]{2011ApJ...737L..22C} or those older than several {tens of thousands of years} \citep[e.g., W44 at $\sim$20,000 yr,][]{1991ApJ...372L..99W}. In the present article, we focus on the young SNRs RX~J1713.7$-$3946 (hereafter RXJ1713) and RCW~86, which are characterized by the {emission of} bright X-rays and TeV $\gamma$-rays. 

The origin of $\gamma$-rays {from} young SNRs is {currently} under debate, and two mechanisms have been {suggested to explain them}. One is {a} leptonic process {in which} cosmic-ray electrons collide with low-energy photons{, boosting} them {into the} $\gamma$-ray {regime} via the inverse Compton process. The other is {a} hadronic process {in which} cosmic-ray protons collide with interstellar protons to produce neutral pions that decay into two $\gamma$-rays. It is important to verify {whether the $\gamma$-rays are of} hadronic origin {for establishing whether} SNRs {are a} major source of cosmic rays in the Galaxy.

In previous {works that considered the hadronic model for} the $\gamma$-ray emissivity of {an} SNR, the number density of targeted interstellar {protons} has been used (e.g., to estimate the cosmic-ray energy from the $\gamma$-ray luminosity). However, strongly shocked matter (e.g., shock-ionized plasma traced by optical {lines} and free--free radio/X-ray continuum and shocked molecular hydrogen observed as near-infrared {lines}), which occupies {only} a small portion of the interstellar medium (ISM), {has mainly been considered as the target gas. Neutral} molecular {and} atomic hydrogen gas {has} not {been} considered to play a major role {in controlling} the {radiative} properties of SNRs, except for {some} pioneering works \citep[e.g.,][]{1994A&A...285..645A}, while the {interaction of} SNRs {with the ISM has been} a subject of interest in some {other} works \citep[e.g.,][]{1988A&AS...75..363D,1998AJ....116..813D,1999AJ....118..930D,2002AJ....123..337D,2002A&A...387.1047D,2004A&A...426..201D,1987A&A...184..279T,1990ApJ...351..157T,1990A&A...237..189T,1998ApJ...505..286S,2004AJ....127.1098S,1999PASJ...51L...7A}. In this article, we focus on the bulk neutral gas {that} dominates the mass {the ISM. Recent theoretical} studies {have revealed} that the associated ISM may play a major role in determining the {radiative} properties of SNRs, and {clarifying} clarify the role of {inhomogeneities in} the ISM in regulating the high-energy radiations {is important \cite[e.g.,][]{2012ApJ...744...71I,2019MNRAS.487.3199C}}. Here we review {methods used} to identify the ISM associated with a SNR, {a} theoretical model {that incorporates} shock--cloud interactions, and {their theoretical} implications {for} the origin of very-high-energy $\gamma$-rays, with a focus on {the} SNRs RXJ1713 and RCW~86.

\section{{The Interstellar Medium (ISM)}}\label{s:ism}
{The} neutral ISM consists of {the} warm neutral medium (WNM) and {the} cool neutral medium (CNM) \citep[e.g.,][]{2011piim.book.....D}. The WNM has {a} density of 0.1--10 cm$^{-3}$ and {a} spin temperature of 300--10000 K, {while} the CNM has {a} density of 10--10$^3$ cm$^{-3}$ and {a} spin temperature of 10--300 K \citep[e.g.,][]{1980ARA&A..18..219M,2003ApJ...586.1067H}. The masses of the two phases are comparable. The CNM {comprises} clumps with a volume filling factor of a few percent, {while} the WNM occupies a large volume of inter clump space \citep[][]{2011piim.book.....D,2018ApJ...860...33F}. The CO clouds are formed in the CNM clumps, and {their} volume filling factor{s are} also very small \citep[$\sim$3--4\%, e.g.,][]{2012ApJ...759...35I,2018arXiv181102224T}. {This picture is significantly different from the conventional assumptions made for (one-dimensional) models, which do not appropriately represent these two-phase inhomogeneities.}

The 21~cm H{\sc i} emission and 2.6~mm CO emission are good tracers of the ISM associated with a SNR. {Comparing a} high-resolution X-ray image with {the} CO and H{\sc i} distributions {provides} a powerful method {for identifying} the {interaction between an SNR and the ISM}. Because electrons cannot penetrate the dense ISM clumps, {we find a small-scale} anti correlation between the X-rays and dense ISM clumps {at sub-pc scales}. Calculations of the ISM column density can be {performed} using {a} CO-to-H$_2$ conversion factor \citep[for a review, see][]{2013ARA&A..51..207B} and the H{\sc i} 21 cm emission, where {corrections} for the H{\sc i} optical depth {are} required for {the} dense CNM.

\begin{figure}[t]
\includegraphics[width=\linewidth,clip]{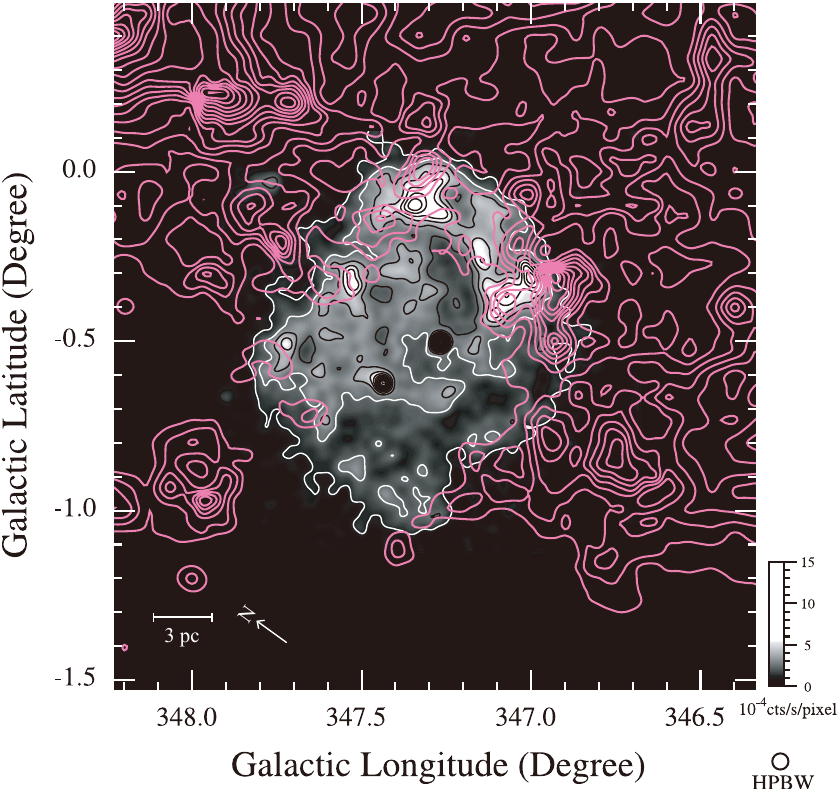}
\caption{Map of {the} {\it{ROSAT}} X-ray flux superposed on the NANTEN2 CO($J$ = 1--0) intensity contours toward RX~J1713.7$-$3946. The CO velocity range is {integrated} from $-11$ to $-3$ km s$^{-1}${, which corresponds} to {the} velocity component interacting with the {supernova remnant (}SNR{)}. The lowest contour level and {contour} intervals are 4 K km s$^{-1}$. From \cite{2003PASJ...55L..61F} with permission}
\label{fig:1}
\end{figure}

{Although} the scheme {described above} works successfully in general, {one} caveat is that the intermediate density regime between 100 and 1000 cm$^{-3}$ may not be detectable in the CO emission {because of the low CO} abundance {caused} by photo dissociation. In addition, the H{\sc i} emission may become optically thick in {this density} regime, requiring a correction for saturation. H{\sc i} is observed {to have} self-absorption dips{; thus,} some H{\sc i} gas is optically thick \citep[e.g.,][]{1978AJ.....83.1607S}.

\begin{figure*}[t]
\includegraphics[width=\linewidth,clip]{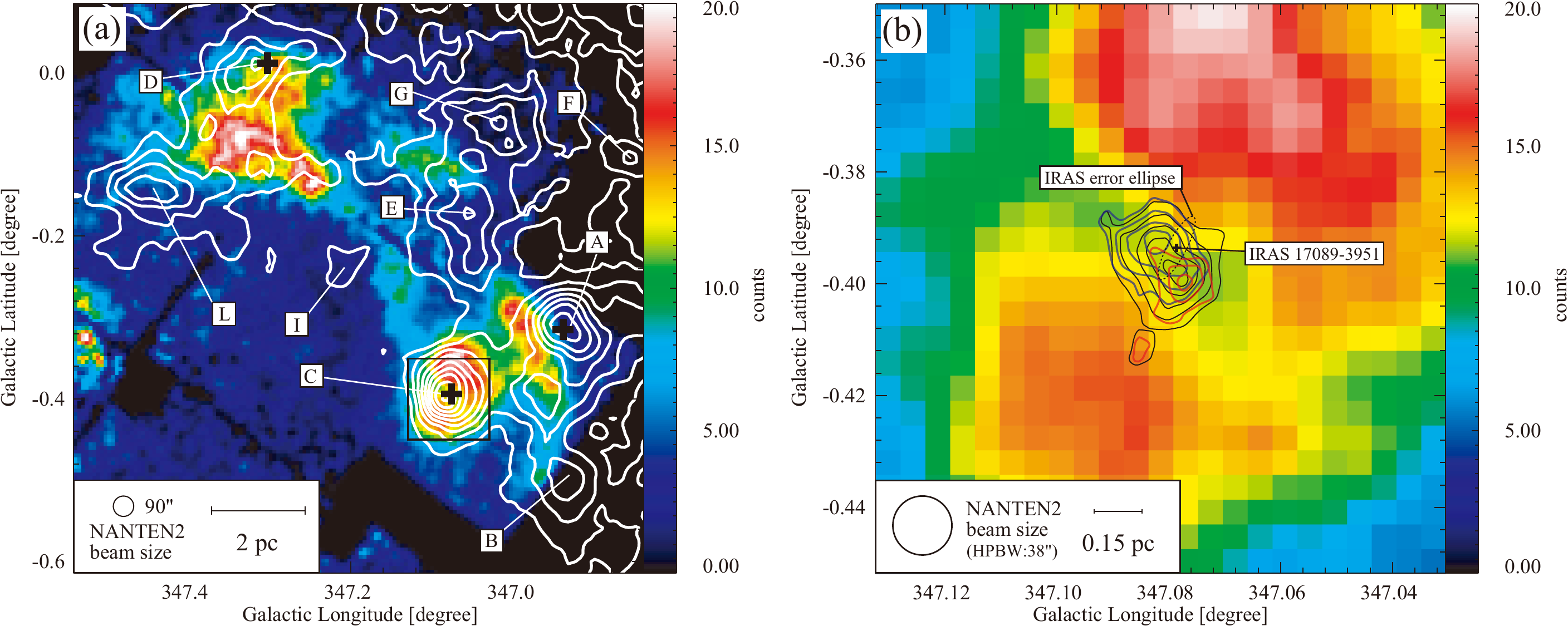}
\caption{(a) {\it{Suzaku}} {5--10 keV}  X-ray image toward the northwestern shell of RX~J1713.7$-$3946. {The} superposed contours {represents the} $^{12}$CO($J$ = 2--1) integrated intensity {obtained with} NANTEN2. The {black} crosses indicate the positions of {\it{IRAS}} {(Infrared Astronomical Satellite)} point sources. The area enclosed by the black {box} is shown enlarged in Figure \ref{fig:2}(b). (b) Enlarged view of {the} $Suzaku$ X-ray image toward CO peak C superposed on the $^{12}$CO($J$ = 4--3) core. {The} black contours {indicate} the total integrated intensity. {The} blue and red contours {represent} the {blue- and red-shifted} outflow components, respectively {(from} \citealt{2010ApJ...724...59S}, reproduced by permission of the AAS)}
\label{fig:2}
\end{figure*}

The existence of dark gas, which cannot be probed by either CO or H{\sc i} emission, was {suggested} {by} {\it{EGRET}} $\gamma$-ray observations \citep{2005Sci...307.1292G} as well as {from} submillimeter observations by {\it {Planck}} \citep[e.g.,][]{2011A&A...536A..19P}. {Two possible explanations exist for} the physical entity of the dark gas{:} CO-dark H$_\mathrm{2}$ gas and optically thick H{\sc i} gas. The {concept of} CO-dark H$_\mathrm{2}$ gas is based on numerical simulations of the evolution of the H$_\mathrm{2}$/H{\sc i} gas {conducted} by \cite{2010ApJ...716.1191W}, who showed that {the} self-shielding of H$_\mathrm{2}$ against UV is strong and a phase of H$_\mathrm{2}$ without {an} observable abundance of CO can {exist} outside {a} CO cloud. This gas, {which comprises} H$_\mathrm{2}$ without CO, is a reasonable model provided that H{\sc i} is converted mainly into H$_\mathrm{2}$. {However,} \cite{2014ApJ...796...59F,2015ApJ...798....6F} used the {\it{Planck/IRAS}} dust emission as {a} proxy {for the} hydrogen column density and found that the H{\sc i} emission becomes optically thick under the usual H{\sc i} gas condition, {in which} the H{\sc i} column density is assumed to be proportional to the dust optical depth. The saturated H{\sc i} intensity becomes weaker than {in} the optically thin case, resulting {in a smaller} H{\sc i} column density {than} under the conventional optically thin approximation. The column density corrected for the optical depth becomes larger than the optically thin H{\sc i} column density {by a factor of 1.5--2;} this increased column density is large enough to explain the dark gas \citep{2017ApJ...838..132O,2019ApJ...878..131H,2019ApJ...884..130H,2020ApJ...890..120M}. The optically thick H{\sc i} raises {the} possibility {of} H{\sc i} {being} dominant compared with H$_\mathrm{2}$, which requires theoretical justification. Simulations by \cite{2012ApJ...759...35I} {using a} gas-evolution code {similar to that employed by} \cite{2010ApJ...716.1191W} presented another {evolutionary} model of H{\sc i} flows {showing} H{\sc i} dominant outside {the} CO clouds. An important {difference exists} between the {initial conditions in these} two simulations. \cite{2012ApJ...759...35I} adopted {a} lower H{\sc i} density{, which includes both} the CNM and WNM. This is {a} realistic H{\sc i} distribution compared with dense H{\sc i} {comprising} only the CNM, which was the initial condition assumed by \cite{2010ApJ...716.1191W}. The lower density  leads to slower H{\sc i}--H$_\mathrm{2}$ conversion than the high-density initial condition and {produces} a rich H{\sc i} envelope surrounding the CO gas instead of CO-dark H$_\mathrm{2}$ gas. This {lends theoretical} support {to} optically thick H{\sc i} as an alternative. A possible new probe {for this} density regime is C{\sc i} emission at {submillimeter wavelengths, which can} serve as another useful tool \citep{2018arXiv181102224T}.

\section{RX~J1713.7$-$3946}\label{s:rxj1713}
\subsection{Distance determination{: non}thermal X-rays and the interacting ISM}\label{ss:distance}
Except for several nearby SNRs with historical records, {determining distances} to SNRs in the Milky Way {is not straightforward,} {because most of SNRs} {lie} in the {Galactic} plane {and are heavily obscured.} Although the relation between radio surface brightness and angular diameters ($\Sigma$--D relation) {has been} used {to estimate} the distances to SNRs, the $\Sigma$--D relation does not provide an accurate measure of distance {because of} its large {scatter} \citep[e.g.,][]{2014SerAJ.189...25P}. {Shock}-excited {masers}---e.g., OH (1720 MHz) masers---{are} very useful {for identifying} the shocked interface, and the {distance to} SNRs {can be obtained from} the radial velocities of the masers, although {this method} can only be used for middle-aged SNRs {that} have interacted {with the ISM} for a long time \citep[e.g.,][]{2013SSPMA..43....1C}. The ISM, including CO clouds and H{\sc i} gas, associated with {an} SNR also provides {a} kinematic velocity, {using} which {the} distance can be calculated by {comparison with a} kinematic model of the Galaxy. The uncertainty is mainly due to the random cloud {motions, which are on} the order of 10 km s$^{-1}$.

\begin{figure}[t]
\includegraphics[width=\linewidth,clip]{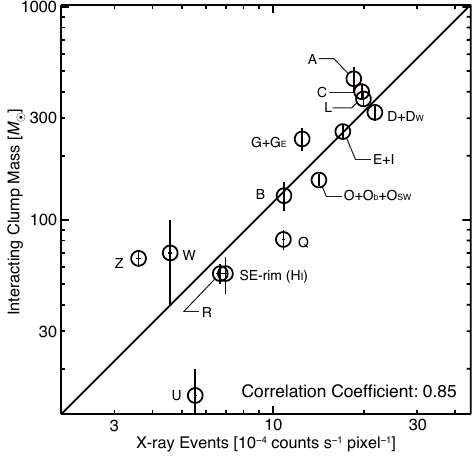}
\caption{Correlation between the peak X-ray flux and the clump mass interacting with the SNR shock. The linear regression {obtained} by least-squares fitting is shown by the solid line, {for which} the correlation coefficient is $\sim$0.85 {on the} double{-logarithmic} scale. {The labels indicate {the} clump name: A, B, C, D$+$D$_\mathrm{W}$, E$+$I, G$+$G$_\mathrm{E}$, L, O$+$O$_\mathrm{b}$$+$O$_\mathrm{SW}$, R, U, W, {and} Z for CO clumps and SE-rim for an H{\sc i} clump.} From \cite{2013ApJ...778...59S}, reproduced by permission of the AAS}
\label{fig:3}
\end{figure}

\begin{figure}[t]
\begin{center}
\includegraphics[width=\linewidth,clip]{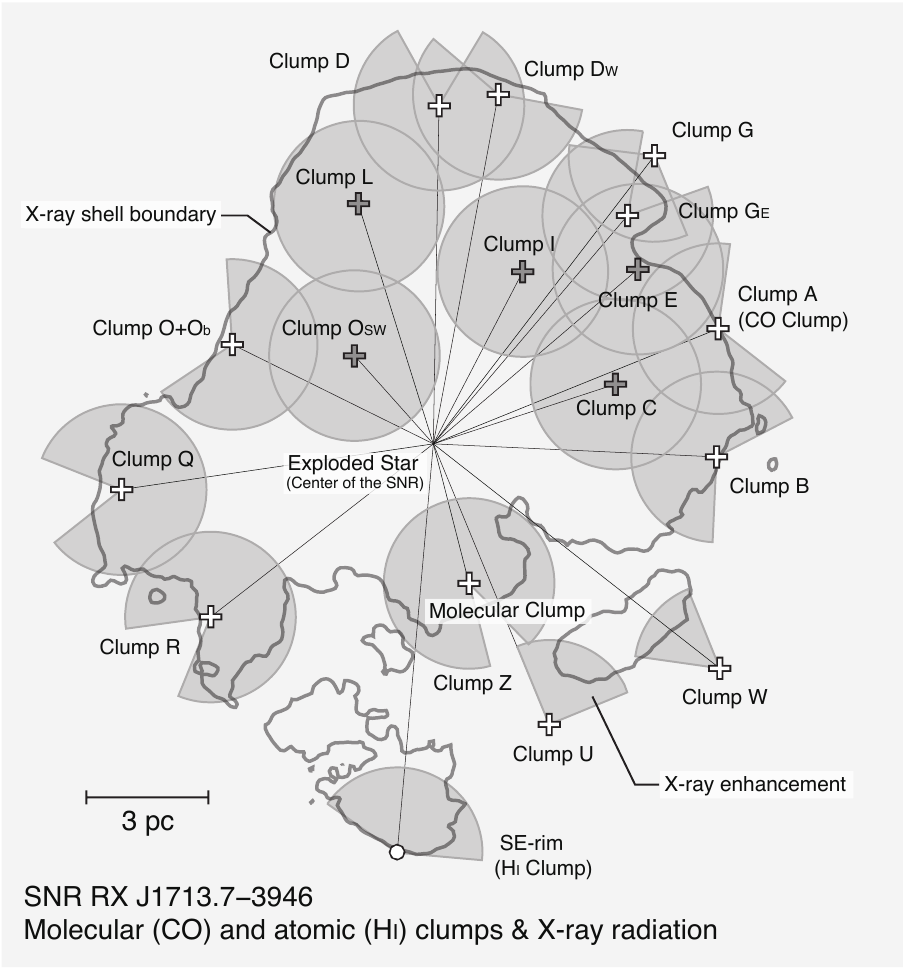}
\end{center}
\caption{Schematic illustration of {the} distributions of molecular (CO) clumps (open crosses), an atomic (H{\sc i}) clump ({open} circle), and X-rays (shaded partial or full circles) superposed on the SNR shell boundary {from the} $Suzaku$ X-rays (gray contours). The black open crosses (CO clumps C, E, I,  L, and O$_\mathrm{SW}$) indicate those fully surrounded by X-rays. From \cite{2013ApJ...778...59S}, reproduced by permission of the AAS}
\label{fig:4}
\end{figure}

\begin{figure*}[h]
\begin{center}
\includegraphics[width=\linewidth,clip]{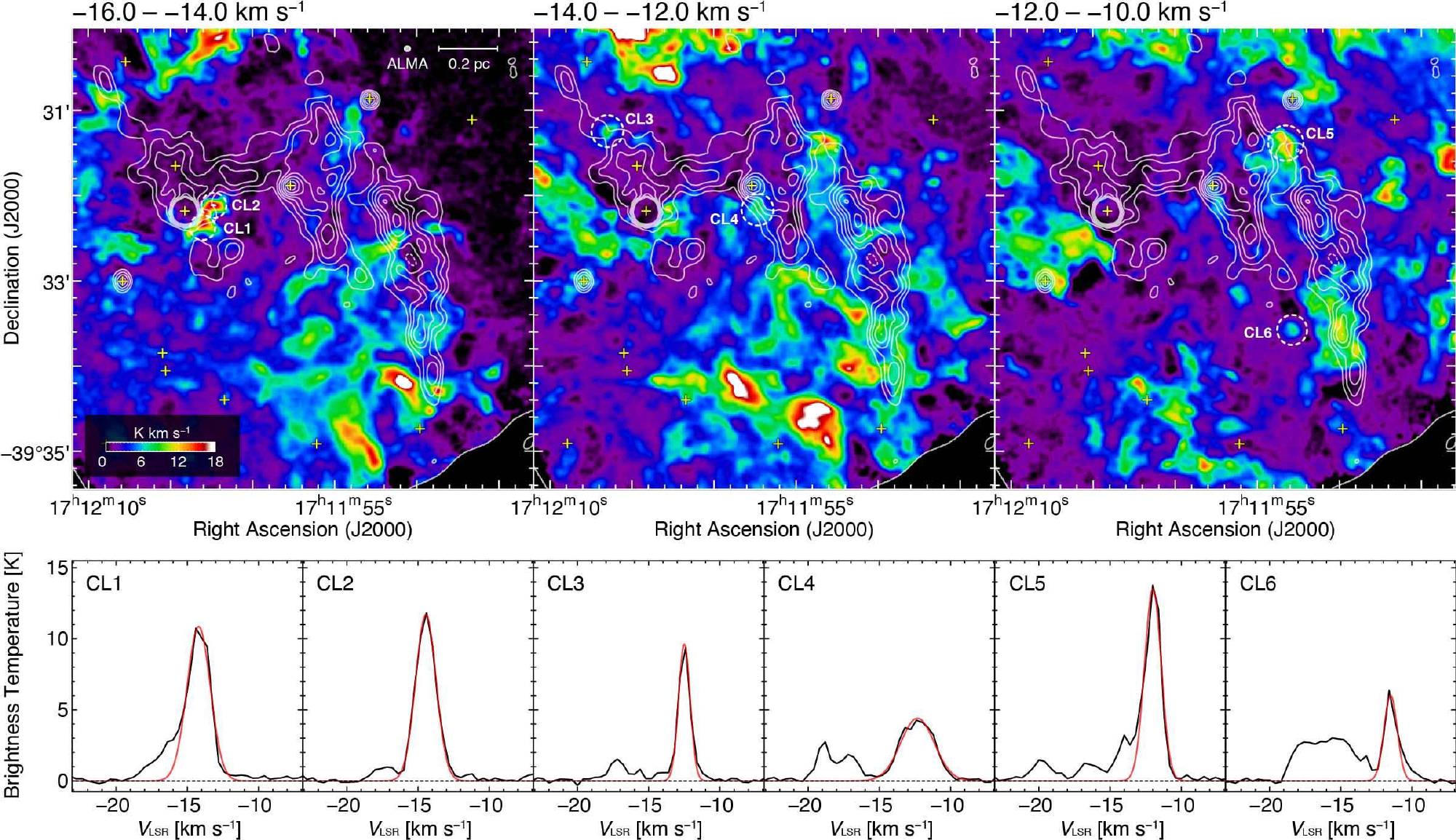}
\end{center}
\vspace*{-0.2cm}
\caption{{\it{Top panels}}: Velocity channel distributions of {Atacama Large Millimeter/submillimeter Array (ALMA)} $^{12}$CO($J$~=~1--0) toward the northwestern shell of RXJ1713. {The superposed} contours {represent} {\it{Chandra}} synchrotron X-ray intensities. {The yellow} crosses represent {the} positions of X-ray hotspots {identified} by \cite{2020ApJ...899..102H}. {Typical} CO cloudlets---named CL1--6---are also indicated. {\it{Bottom panels}}: CO profiles of CL1--6 (black lines) and Gaussian curves (red lines) fitted using the least-squares method. From \cite{2020ApJ...904L..24S}, reproduced by permission of the AAS}
\label{fig:new}
\end{figure*}

The nonthermal X-ray SNR RXJ1713 {is} a case {for which} two distances {that differ} by a factor of 6 {had been} debated. It was first suggested that {this} SNR is located at a distance of 1 kpc, assuming a typical average H{\sc i} density {for} the X-ray absorption \citep{1997PASJ...49L...7K}, and {it} was identified as the SNR {of 393AD from} Chinese historical records \citep{1997A&A...318L..59W}. \cite{1999ApJ...525..357S} claimed that CO clouds {at} $-69$ {and} $-94$ km s$^{-1}$ are associated with the SNR, {from} 8.8 arcmin resolution {maps obtained} with the CfA~1.2-m telescope, and {they} derived a large kinematical distance ($\sim$6 kpc). These authors argued that the H{\sc i} density toward RXJ1713 is lower than the typical Galactic average {owing} to {an} accidental coincidence with a hole {in the} ISM, while {a} three-times-larger distance {implies} an unusually large radius of $\sim$50 pc for an age of $\sim$$10^4$ yrs. The 6 kpc distance {has been} questioned by a subsequent {high-}resolution CO survey, which {found} another CO component at $-10$ km s$^{-1}$ to be associated {with this SNR at the} 2.6 arcmin resolution {CO maps} {obtained} with the NANTEN 4-m telescope, as shown in Figure \ref{fig:1} \citep{2003PASJ...55L..61F,2005ApJ...631..947M}. The observational signature {of this} association is the large-scale correlation ({on scales of} a few pc) between the CO {map} and the {\it{ROSAT}} X-ray {survey}, which was confirmed {in} more detail by the {\it{Suzaku}} data {from} \cite{2010ApJ...724...59S,2013ApJ...778...59S} (see Section \ref{ss:detailed}). {As the} low velocity CO gas is fainter by an order of magnitude than the $-69$ {and} $-94$ km s$^{-1}$ clouds, {it} was {ignored} by \cite{1999ApJ...525..357S}. The {smaller} distance is now widely accepted, and it {has been} established that RXJ1713 is a young SNR {aged} 1600 yr{, which was} historically recorded \citep{2004Natur.432...75A,2004A&A...427..199C,2008ApJ...685..988T,2015ApJ...799..175S,2016PASJ...68..108T}. This shows that high resolution and sensitivity {are} crucial in a comparison with the ISM, raising {questions about} the distance determinations in previous {low-resolution studies}.

\begin{figure}[]
\begin{center}
\includegraphics[width=80mm,clip]{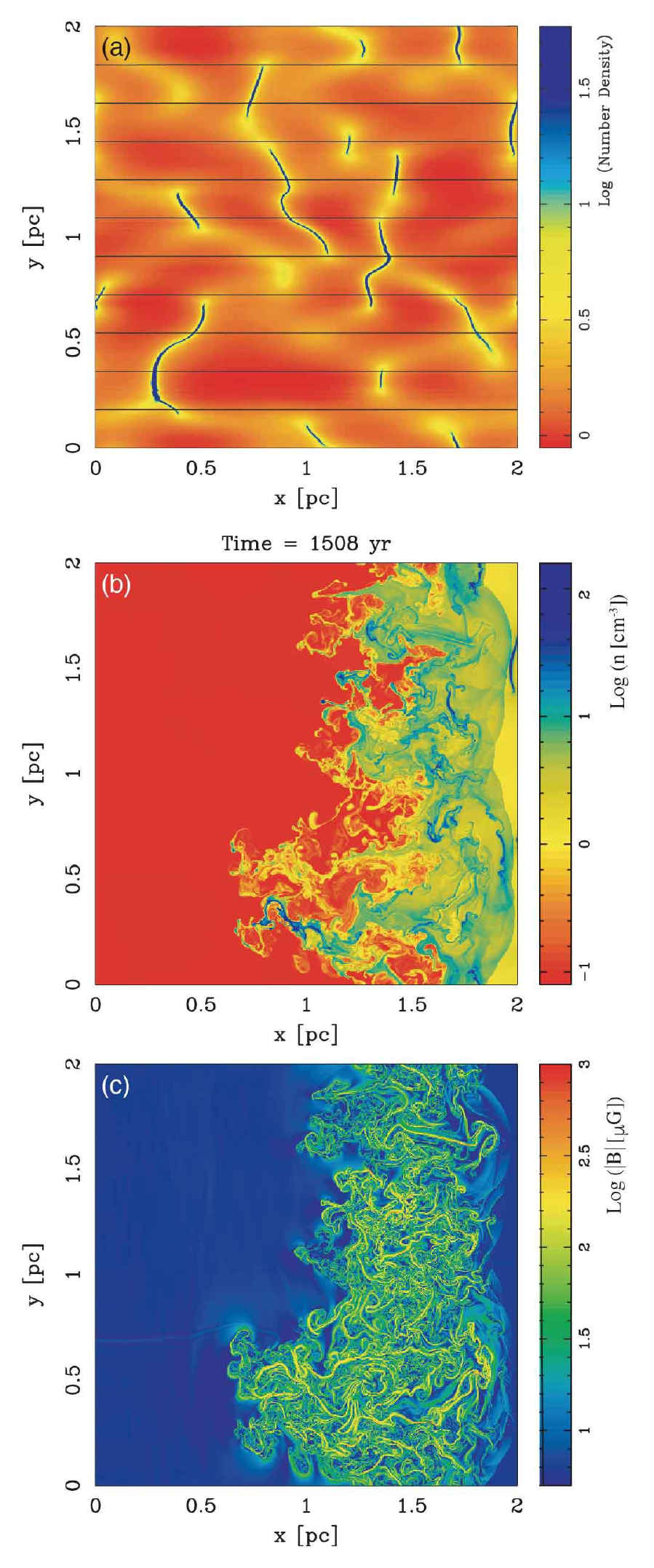}
\end{center}
\caption{(a) {Structure} of {the} interstellar medium generated by the thermal instability. {The} colored image shows {the} number density of atomic hydrogen, and the black lines {represent} magnetic field lines. (b, c) Results {at} $t = 1508$ yr after the injection {of the parallel shock}: (b) number density of atomic hydrogen and (c) the magnetic field strength. From \cite{2009ApJ...695..825I}, reproduced by permission of the AAS}
\label{fig:5}
\end{figure}

\subsection{Detailed correspondence between X-rays and ISM clumps}\label{ss:detailed}
\cite{2012ApJ...746...82F} showed that RXJ1713 is associated with molecular and atomic gas, {each} having masses of $\sim$$10^4$ $M_\odot$. The ISM distribution is shell-like{, which an} 8 pc radius, and {the} molecular cloud cores are inside the shell. The shell is consistent with {a} core-collapse SNR, {for which} the stellar winds {from} the high-mass progenitor evacuated the {surrounding} ISM prior to the SN explosion.

\begin{figure}[h]
\begin{center}
\includegraphics[width=\linewidth,clip]{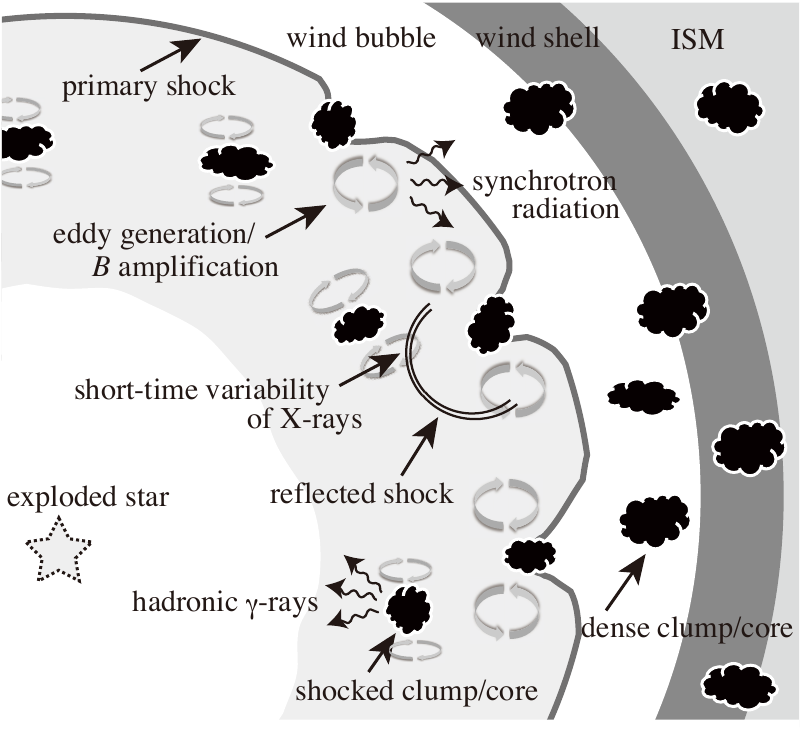}
\end{center}
\vspace*{-0.4cm}
\caption{Schematic of the shock--cloud interaction model (see the text). From \cite{2012ApJ...744...71I}, reproduced by permission of the AAS}
\label{fig:6}
\end{figure}

A comparison of the {\it Suzaku} X-ray data with the CO image revealed details of the interaction \citep{2010ApJ...724...59S}. Figure \ref{fig:2} shows that the CO peaks {A--E, G, I, and L} are anticorrelated with the X-ray peaks\footnote{Note that these anti-correlations are not due to interstellar absorption, because the 5--10 keV X-ray energy band is thought to be an absorption-free energy range when the ISM proton column density is {less} than 10$^{23}$ cm$^{-2}$.}, indicating that the dense cores hinder cosmic-ray electrons {from penetrating} them, and the X-rays are rim-brightened on the surfaces of the CO cores {\citep[see also][]{2012MNRAS.422.2230M}.} The cores in the shell were formed via gravitational instability over a timescale of Myr prior to the SN explosion, and three of them {(CO peaks A, C, and D)} are forming young stars, as shown by the infrared point sources \citep{2010ApJ...724...59S}. {In fact, peak C shows obvious signs of active star formation, {including} bipolar outflows and an IRAS point source (see Figure \ref{fig:2}a).} 

\cite{2013ApJ...778...59S} {conducted} a comprehensive study of the dense cores probed by CO as well as {by} cold H{\sc i}, and found that the X-ray {intensities are} well correlated with the masses of the ISM cores, as shown in Figure \ref{fig:3}. \cite{2013ApJ...778...59S} {showed} that more than 80\% of the X-ray peaks, brighter than $5 \times 10^{-4}$ counts s$^{-1}$ pixel$^{-1}$, are associated with the ISM cores, suggesting {a tight close} connection between X-ray {emission} and the {masses of the} interacting ISM cores (Figure \ref{fig:4}). This suggests that the distribution of the dense gas {had a significant effect on} regulating the X-ray distribution in the recent 1000 yr{, assuming a} shock velocity of {$\sim$4000} km s$^{-1}$ {\citep[c.f.,][]{2016PASJ...68..108T}}, {which has} stimulated a detailed theoretical study {of} the interaction between {a} shock front and clumpy ISM{\footnote{{According to proper-motion measurements of {the} {\it{Chandra}} X-rays, the shock speed is measured to {be} $3900 \pm 300$ km s$^{-1}$ \citep{2016PASJ...68..108T}. The shock propagates through low-density intercloud regions without much deceleration (see also Section \ref{ss:simulations}). {Because} the shocked molecular clouds are located within $\sim$4 pc from the shock front, the clouds {have} regulated the X-ray distribution in the {most} recent 1000 yr.}}}. 

Most recently, \cite{2020ApJ...904L..24S} {obtained} high-resolution CO observations ({with} $\sim$0.02~pc {resolution}) toward the northwestern shell of RXJ1713 using the Atacama Large Millimeter/submillimeter Array (ALMA). The authors {found dozens of} tiny clumps, with typical radii of $\sim$0.03--0.05~pc ({referred} to as {\it{molecular cloudlets}}). Figure \ref{fig:new} shows {the} velocity channel distributions of ALMA CO {intensities} superposed on {\it{Chandra}} synchrotron X-ray contours. The molecular cloudlets are located not only along {the} synchrotron X-ray filaments, but also in the vicinity of X-ray hotspots {that exhibit flux variations on timescales} of a few months or years \citep[see][]{2007Natur.449..576U,2020ApJ...899..102H}. {Because the} CO profiles of {the} cloudlets {have} narrow widths ({see the bottom panels in} Figure \ref{fig:new} ), the authors concluded that these clumpy structures were formed before the supernova explosion.

\subsection{Simulations of shock-cloud interactions}\label{ss:simulations}
Magneto-hydrodynamical (MHD) numerical simulations of {a} supernova shock propagating {in a} clumpy ISM show that {both} the turbulence and the magnetic field are amplified around {the} dense cores, offering a theoretical basis for the ISM--X-ray correspondence \citep{2009ApJ...695..825I,2012ApJ...744...71I}.

Figures \ref{fig:5}(a) and \ref{fig:5}(b) show the density distributions at 0 and 1508 yr {after} since the onset of the shock interaction. The low-density gas {(with number density of $\sim$}1 cm$^{-3}${)---i.e.,} the WNM{---}is disturbed significantly by the shock acceleration, while the high-density gas {(}with {number densities exceeding} 100 cm$^{-3}${)---i.e.,} the CNM{---}is not much accelerated. In spite of the interaction, the shock front propagates almost freely at {its} initial velocity {of} 3000 km s$^{-1}$ {owing} to {the} low filling factor of the CNM cores. Figure \ref{fig:5}(c) shows the distribution of the magnetic field which indicates {a} highly entangled field configuration. The shock front, which was initially {planar,} is deformed when it hits a dense core and becomes entangled around the core. This {produces a} turbulent velocity field in the WNM, and the magnetic field becomes turbulent and {is} amplified to 100 $\mu$G or higher from its initial value {of} 6 $\mu$G \citep{2009ApJ...695..825I}. The amplified magnetic field surrounds the dense cores, as {shown} in Figure \ref{fig:5}(b), which explains {why} the synchrotron X-rays {are} rim-brightened around the dense cores. The simulations show that the magnetic field is amplified around the dense core {in a layer of} {sub} pc-scale thickness, where the nonthermal X-rays are enhanced.

\begin{figure}[]
\begin{center}
\includegraphics[width=\linewidth,clip]{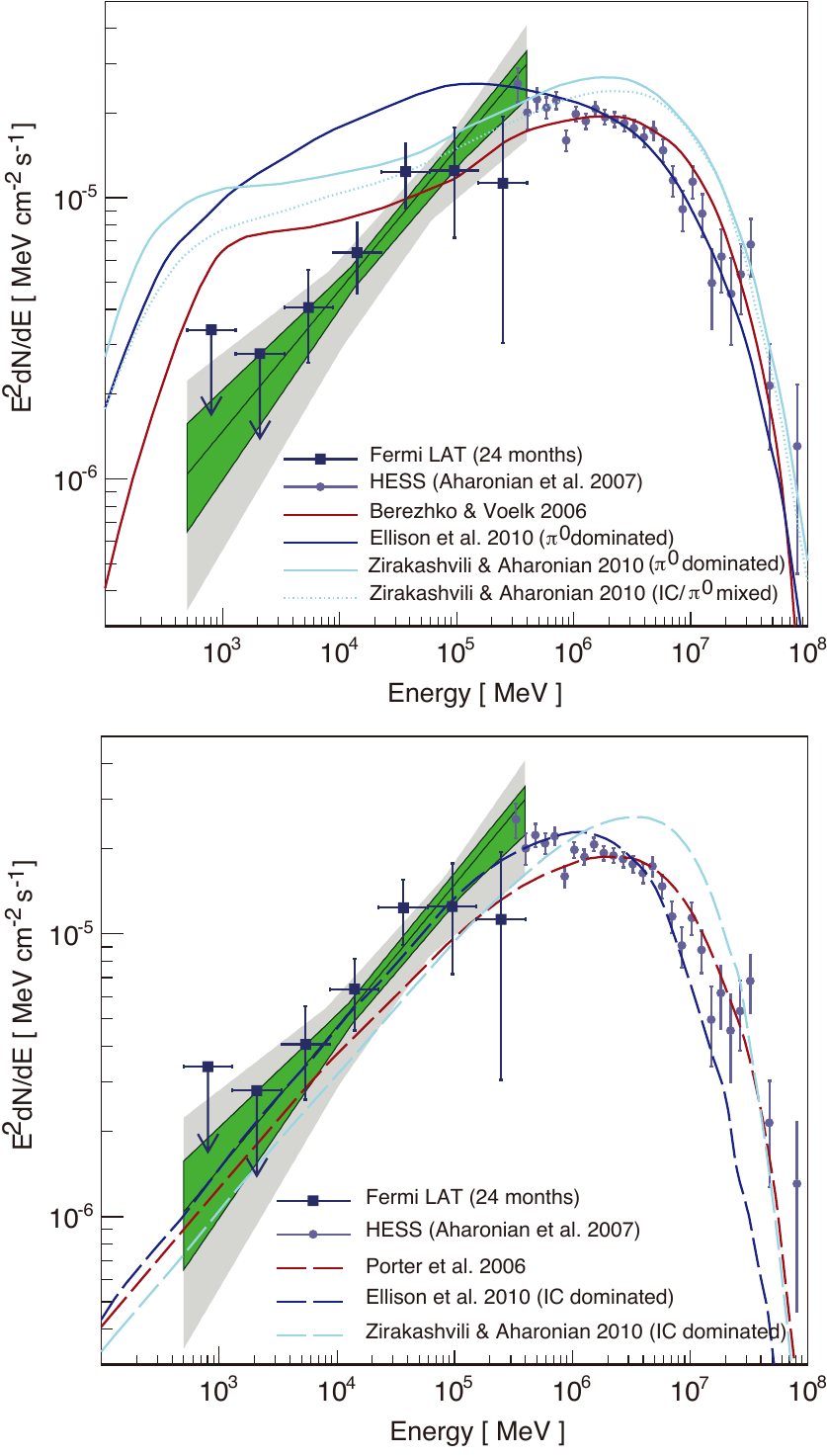}
\end{center}
\caption{{Spectral} energy distributions (SEDs) of RX~J1713.7$-$3946 in $\gamma$-rays. The green-shaded areas show the uncertainty bands obtained from maximum-likelihood fits, assuming a power-law spectrum {extending} from {0.5} to 400 GeV. The gray-shaded areas represent systematic {uncertainties in} the model {fits. The solid} and dashed curves {represent various} models \citep{2006A&A...451..981B,2006ApJ...648L..29P,2010ApJ...712..287E,2010ApJ...708..965Z}. The upper panel {shows} the hadronic models, whereas the bottom panel {displays} the leptonic models. From \cite{2011ApJ...734...28A}, reproduced by permission of the AAS}
\label{fig:7}
\end{figure}

Figure \ref{fig:6} shows a schematic of the shock--cloud interaction model {of} \cite{2012ApJ...744...71I}. Before the supernova explosion of a high-mass progenitor, ambient interstellar gas such as diffuse H{\sc i} is completely evacuated by {a} strong stellar wind. The wind creates a wind bubble {with a} density {of} $\sim$0.01 cm$^{-3}$ \citep[e.g.,][]{1977ApJ...218..377W}. By contrast, dense molecular clouds{, with densities exceeding} $\sim$1000 cm$^{-3}$ can survive in the wind. {Consequently, an} inhomogeneous gas distribution with a density difference of five orders {of} magnitude {is} formed by the high-mass progenitor, {before} the supernova explosion {occurrs}. The primary forward shocks {from the SN explosion} propagate through the wind bubble with clumpy clouds, where particle acceleration operates. {Shocks transmitted through} the gas cloud are stalled, whereas shock waves in the intercloud {medium} are not decelerated. These velocity differences are also observed {in measurements of the} proper motions of {the} forward shocks \cite[e.g.,][]{2016PASJ...68..108T}. The velocity difference generates multiple eddies around the shocked clouds, which enhance the magnetic field strength up to $\sim$1 mG. Finally, we observe that the shocked clouds are limb-brightened by the synchrotron X-rays. The high magnetic field strength causes short{-timescale} variability of {the} synchrotron X-rays, as discovered by \cite{2007Natur.449..576U}. The shocked gas clouds also act as {targets for} cosmic-ray protons and produce hadronic $\gamma$-rays (see Sections \ref{ss:gspectra} and \ref{ss:ism-gamma}).

\subsection{The clumpy gas distribution and the $\gamma$-ray spectrum}\label{ss:gspectra}
Many previous theoretical works {on gamma-ray production via the hadronic process have} assumed uniform {density} or radial density gradients, with {a} typical ambient density {of $\sim$}1 cm$^{-3}$ \citep[e.g.,][]{2010ApJ...712..287E,2013ApJ...767...20L}. Even though {a} low-density environment such as a wind-swept cavity and non-uniformity {have been} discussed to explain the observed $\gamma$-ray spectra, potentially energy dependent propagation effects in the multi phase ISM {have} not {been} considered \citep[cf.,][]{2010A&A...511A..34B}. One {reason} why the multi phase ISM {has} not generally {been} considered is {because} MHD effects {in} SNRs are mostly negligible for low-density contrasts $\la10^2$ \citep[][]{2013ApJ...763...14B}. The real ISM environment surrounding a SNR, however, {contains} preexisting inhomogeneities {with density contrasts of} $\sim$10$^5$ in the ISM, the importance of which was first emphasized by \cite{1974ApJ...189L.105C} and \cite{1977ApJ...218..148M}.

{Conventional} $\gamma$-ray spectra in the hadronic and leptonic scenarios {for a} uniform (or {small density-}contrast) ISM are shown in Figure \ref{fig:7} \citep{2011ApJ...734...28A}. The hadronic spectrum is {relatively} flat, {with {a} spectral {index} $\Gamma$ of $\sim$2.0}, while the leptonic spectrum is hard in the GeV band {($\Gamma {\simeq}1.5$)}. {Therefore, on average, over several SNRs---the shape of the $\gamma$-ray spectra---}should allow {discrimination} between the two scenarios if we consider a simple one-zone assumption. However, such assumptions {do} not hold for individual SNRs{, as} has been discussed in the community, even {for} the typical young SNR Cassiopeia~A \citep[e.g.,][]{2000A&A...354..915A,2008ApJ...686.1094H}. RXJ1713 is also {a} representative examples: the inclusion of the clumpy ISM has a significant effect and significantly changes the hadronic spectrum as discussed below \citep{2010ApJ...708..965Z,2012ApJ...744...71I,2014MNRAS.445L..70G,2019MNRAS.487.3199C}.

The hadronic $\gamma$-ray scenario requires {a} high density of target protons, whereas the large volume {of low-density gas} is required for efficient particle acceleration via diffusive shock acceleration (DSA). Because the maximum energy of {the} accelerated particles is proportional to {the} shock speed, it is {assumed} to be that {a} low-density medium (typically less than $\sim$1 cm$^{-3}$) and the large volume is needed to efficient acceleration. Under {the} assumption of {a} uniform ISM, {an} ISM density {greater} than 1 cm$^{-3}$ is required to produce the $\gamma$-rays {observed from} RXJ1713 \citep{2010ApJ...712..287E}. For {such} {high} {densities}, strong thermal X-rays due to shock heating are expected. This {is contradicted} the purely nonthermal X-rays observed {from} RXJ1713. Although some theoretical studies {have} avoided {this} contradiction by {considering a} wind-bubble model and/or {a} thermal nonequilibrium state between {the} electrons and protons downstream \citep[e.g.,][]{2009MNRAS.392..240M,2010A&A...511A..34B},  the absence of bright thermal X-ray emission {has been} used to exclude hadronic $\gamma$-rays and {support} leptonic $\gamma$-rays \citep{2012ApJ...744...39E}. 

{However, the lack of thermal X-ray emission in the presence of high gas densities can be explained naturally by considering a realistic clumpy ISM distribution.} Most of the volume of the interclump medium is of low density, which allows the DSA to work, whereas a large fraction of the mass of the ISM is contained in dense cores, which can work as {a} dense, massive, target material, {enabling the} cosmic-ray protons {to produce hadronic $\gamma$-rays via} the p--p reaction. In the dense cores, the shock fronts are decelerated {significantly,} with no heating, and thermal X-rays are suppressed, {as indicated by the} equation \ref{eq0} \citep{2012ApJ...744...71I}:
\begin{eqnarray}
k_\mathrm{B} T_\mathrm{c} =  2 \times 10^{-4}\;  \biggl(\frac{v_\mathrm{sh,d}}{3000\;\mathrm{km\;s^{-1}}}\biggr)^2\; \biggl(\frac{n_\mathrm{d}}{0.01\;\mathrm{cm^{-3}}}\biggr)\nonumber\\
\biggl(\frac{n_\mathrm{c}}{10^3\;\mathrm{cm^{-3}}}\biggr)^{-1} \;(\mathrm{keV}),
\label{eq0}
\end{eqnarray}
where $k_\mathrm{B}T_\mathrm{c}$ is the maximum temperature of {the} protons, $v_\mathrm{sh,d}$ is the shock velocity in the intercloud or diffuse region, $n_\mathrm{d}$ is the number density of {the} intercloud or diffuse gas, and $n_\mathrm{c}$ is the number density {in} the dense clouds. {Because} the {density differences} between $n_\mathrm{d}$ and $n_\mathrm{c}$ {are on the} order of 5, no strong thermal X-rays {are} expected {from} RXJ1713 under the clumpy ISM distribution. {The} recently discovered thermal X-ray line emission {from} RXJ1713 originates from {the} supernova ejecta \citep{2015ApJ...814...29K}, and hence it is consistent with the shock--cloud interaction model.

\cite{2012ApJ...744...71I} studied $\gamma$-ray production in the shock--cloud interaction in RXJ1713 based on {MHD} numerical simulations, and {they} showed that a hard spectrum {($\Gamma_\mathrm{GeV} = 1.5$ in the GeV band)} like {that of} leptonic $\gamma$-rays is well reproduced in the hadronic scenario with {a} clumpy ISM. Within the SNR shell, the dense cores are exposed to cosmic-ray protons, and the p--p interactions inside the cores produce $\gamma$-rays. The p--p reactions thus lead to {a} spatial correspondence between the $\gamma$-rays and the ISM mass.

\begin{figure}[t]
\begin{center}
\includegraphics[width=\linewidth,clip]{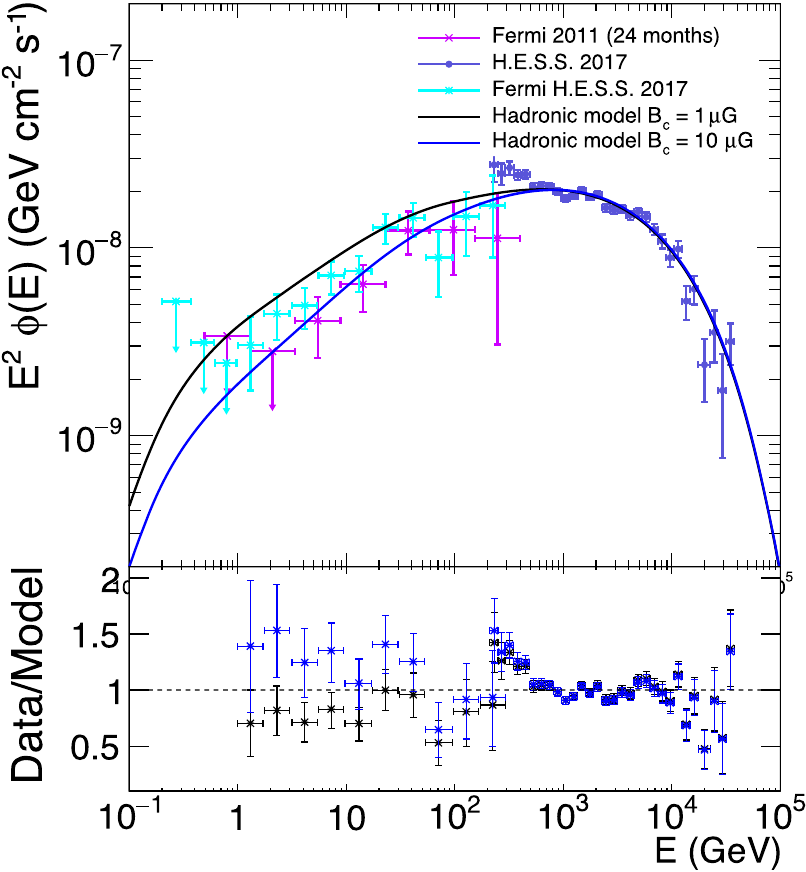}
\end{center}
\caption{Hadronic models for {the} $\gamma$-ray SEDs of RX~J1713.7$-$3946. The hadronic models (solid lines) refer to {a} configuration with a magnetic field strength inside the gas clump {of} $B_\mathrm{c}$ = 1 $\mu$G (black) {or} to $B_\mathrm{c}$ = 10 $\mu$ (blue). The field in the {skin of the} gas clump is fixed {at} $B_\mathrm{s}$ = 100 $\mu$G in both models. The bottom pane shows {the} residuals for the data/model. From \cite{2019MNRAS.487.3199C} with permission}
\label{fig:8}
\end{figure}

\begin{figure*}[h]
\begin{center}
\includegraphics[width=\linewidth,clip]{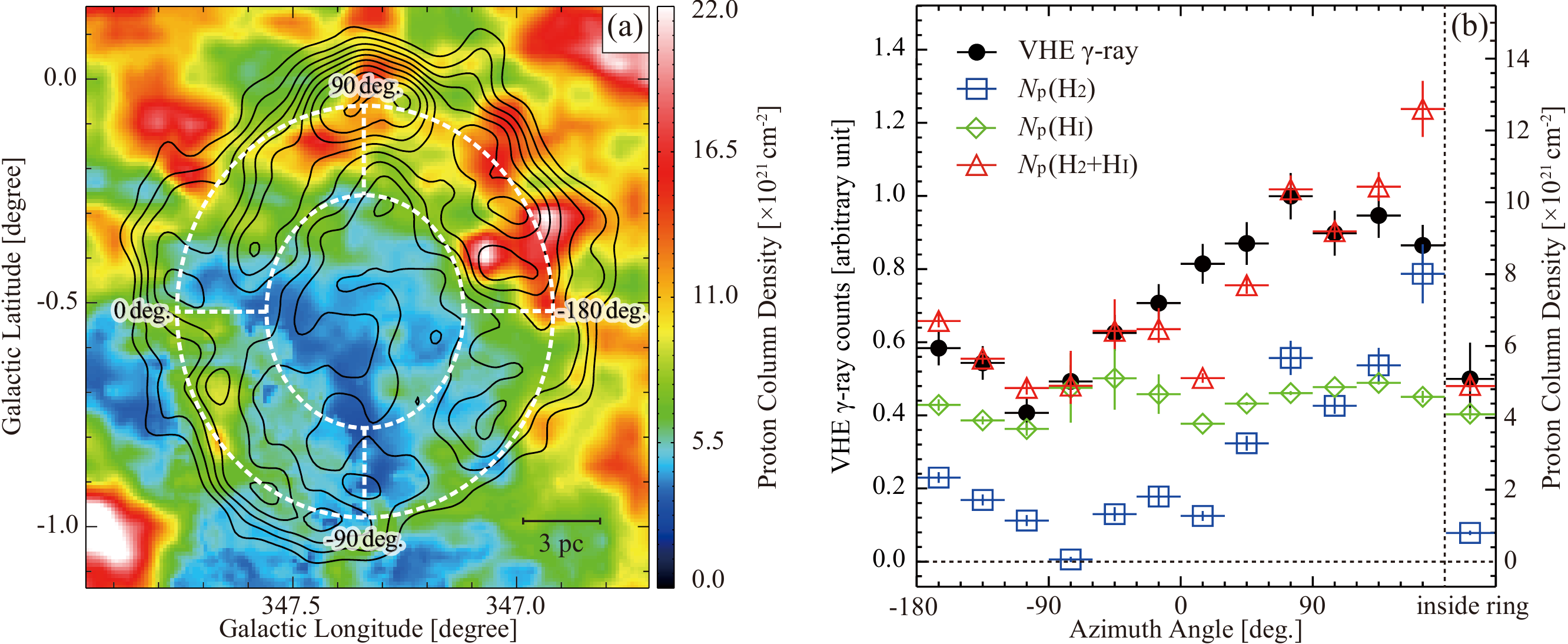}
\end{center}
\caption{(a) Distribution of {the} total proton column density $N_{\mathrm{p}}$(H$_2$+H{\sc i}) {toward} RX~J1713.7$-$3946 superposed on the TeV $\gamma$-ray contours. The lowest contour and {the contour} intervals are 20 and 10 smoothed counts, respectively. (b) Azimuthal distributions of {the} proton column densities of molecular hydrogen $N_{\mathrm{p}}$(H$_2$), atomic hydrogen $N_{\mathrm{p}}$(H{\sc i}) and $N_{\mathrm{p}}$(H$_2$+H{\sc i}), and $\gamma$-rays between the two elliptical rings shown in (a). The same plot {for the area} inside the inner ring {is} also shown on the right-hand side of (b). From \cite{2012ApJ...746...82F}, reproduced by permission of the AAS}
\label{fig:9}
\end{figure*}

{The most essential argument of \cite{2012ApJ...744...71I} is taking into account} the penetration {depths} of cosmic-ray protons into dense cores, which {are} limited by the amplified turbulent magnetic field around {the} dense cores (\citealt{2012ApJ...744...71I}: see also \citealt{2010ApJ...708..965Z}). This tends to dilute the $\gamma$-ray--ISM correspondence. Equation \ref{eq1} gives an expression {for} the penetration depth as a function of the magnetic field {from} \cite{2012ApJ...744...71I},
\begin{eqnarray}
l_\mathrm{pd} =  0.1\; \eta^{0.5}\;  (E / 10\;\mathrm{TeV})^{0.5}\; (B / 100\;\mathrm{\mu G})^{-0.5}\nonumber \\
(t_\mathrm{age} / 1000\;\mathrm{yr})^{0.5} \;\;(\mathrm{pc}),
\label{eq1}
\end{eqnarray}
where $E$ is the particle energy, $t_\mathrm{age}$ is the age of the SNR, and the cosmic-ray energy spectrum is {assumed to have the form} $N(E)dE \propto E^{-p} dE$ above the critical energy for $\pi^0$ creation and below the maximum energy of {the} accelerated nuclei. {Here,} $\eta$ is the degree of randomness of the turbulent magnetic field, {which} is $\sim$1 around the cores \cite[e.g.,][]{2007Natur.449..576U}. This indicates that the cosmic-ray protons can penetrate {on the order of 0.1 pc} into the surface layers of the cores in; {this} is smaller than the typical size {of 1 pc} of the CO cores in RXJ1713. {Consequently}, low-energy cosmic-ray protons cannot penetrate deeply into the dense regions of the cores, and the $\gamma$-ray spectrum is modified {accordingly} \citep{2012ApJ...744...71I,2014MNRAS.445L..70G,2019MNRAS.487.3199C}. The ISM {mass that} interacts with the cosmic rays is proportional to {the} cosmic-ray energy $E$: $M(E) \propto l_\mathrm{pd}(E) \propto E^{1/2}$, if we assume {a} cosmic-ray distribution with the spectral index $p = 2$ {for} high-energy nuclei{, as} in conventional DSA theory \citep{2012ApJ...744...71I}. The increase {in} $M$ with $E$ {produces} a hard $\gamma$-ray spectrum ($\Gamma {\approx} 1.5$), similar to the leptonic spectrum {produced} in {a} uniform low-density ISM. Figure \ref{fig:8} shows {the} $\gamma$-ray SED of RXJ1713, which can be well reproduced by the hadronic scenario {when} the penetration depths of {the} cosmic-ray protons {are considered} \citep{2019MNRAS.487.3199C}. In addition, the penetration depth {is expected to cause} local anticorrelations between the $\gamma$-rays and {the} ISM density {on the} 0.1-pc scale{s}, which may become directly observable in {the} future {using} high-resolution $\gamma$-ray observations with the Cherenkov Telescope Array (CTA).

\subsection{Correspondence between the ISM and $\gamma$-rays}\label{ss:ism-gamma}
\subsubsection{ISM--$\gamma$-ray correspondence}\label{sss:correspondence}
In early works \citep{2006A&A...449..223A,2008AIPC.1085..104F}, only CO clouds were {compared} with the $\gamma$-ray {distributions}. \cite{2012ApJ...746...82F} made a comprehensive comparison between the ISM ({including} both molecular and atomic hydrogen) and {the} TeV $\gamma$-rays observed {using} H.E.S.S. aiming to test the spatial correspondence. A new feature {of this work} was that the H{\sc i} emission was included in the comparison, and the amount of the H{\sc i} gas was found to be comparable {with} the molecular mass. Figure \ref{fig:9} shows that the azimuthal distributions of the two quantities have good spatial correspondence at an effective resolution of 3 pc, which is limited by the H.E.S.S. observations. {These} results lend support {to} the hadronic scenario (see below for further discussion). The 3-pc resolution is {considerably} larger than the typical penetration depth {of $\sim$}0.1 pc and the effect of the {exclusion of} low-energy cosmic rays from the dense region is not significant in the plot. The {good spatial correspondence between the total ISM protons and $\gamma$-rays} {shows for the first time} that the necessary condition for the hadronic origin of the $\gamma$-rays, which was not clear in \cite{2006A&A...449..223A}, is satisfied. Subsequent detailed comparative studies among the ISM, gamma-rays, and synchrotron X-rays supported this result by separating the hadronic and leptonic gamma-rays quantitively \citep{2021arXiv210502734F}.

\begin{figure*}[h]
\begin{center}
\includegraphics[width=\linewidth,clip]{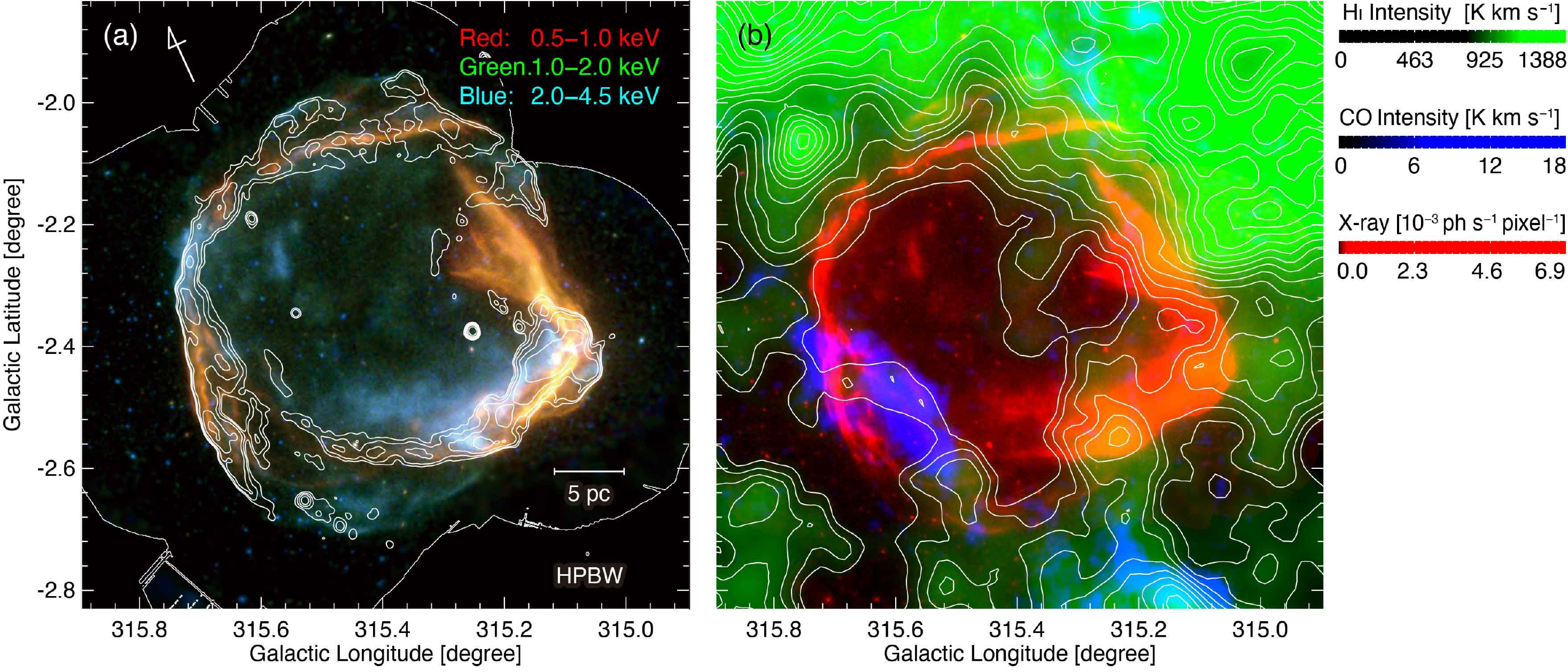}
\end{center}
\caption{(a) RGB image of RCW~86 observed by $XMM$-$Newton$. {Red}, green, and blue represent the X-ray energy bands, 0.5--1.0, 1.0--2.0, and 2.0--4.5 keV, respectively. The white solid lines {outline} the region observed. {The} contours represent the radio continuum {centered} at a frequency 843 MHz. (b) RGB image of RCW~86 and its surroundings. {Here,} red, blue, and green represent the {{\it XMM-Newton}} X-rays (0.5--4.5 keV), NANTEN2 $^{12}$CO($J$ = 2--1), and the ATCA \& Parkes H{\sc i} {map}, respectively. The velocity ranges of CO and H{\sc i} are from $-46.0$ to $-28.0$ km s$^{-1}$. The contours indicate the H{\sc i} integrated intensity. From \cite{2017JHEAp..15....1S} with permission}
\label{fig:10}
\end{figure*}

\subsubsection{Cosmic-ray energy}\label{sss:cr-energy}
The total cosmic-ray energy $W_\mathrm{tot}$ is estimated {from} equation \ref{eq2} {below} \citep[from ][]{2006A&A...449..223A} to be $\sim$$10^{48}$ erg, {which corresponds} to 0.1\% of the total energy of a typical supernova explosion.
\begin{eqnarray}
{W_\mathrm{tot} \sim 1-3 \times 10^{50} \biggl(\frac{d}{1\, \mathrm{kpc}} \biggr)^2 \biggl(\frac{n}{1\, \mathrm{cm^{-3}}}\biggr )^{-1}} \ \ {\rm (erg),}
\label{eq2}
\end{eqnarray}
where $d$ is the {1-kpc} distance to the source and $n$ is the average density of interstellar protons associated with the SNR. {T}his value {is unlikely to represent} the total cosmic-ray energy involved in RXJ1713. The number of dense cores in the shell is {around} 20, {each with} a typical radius of 1 pc \citep{2013ApJ...778...59S}. For the volume of the shell{, which has an} 8-pc radius, the volume filling factor of the cores is estimated to be $\sim$10\%, implying that about {1/10} of the cosmic-ray protons are interacting with the ISM protons. This means that the total energy of cosmic rays is {10} times larger than the estimated value of $W_\mathrm{tot}$. The time evolution {must} be taken into account further, and we expect the {maximum energy} of {the} $\gamma$-rays {to shift} to lower {values} with {10-fold} increase in {the} total energy. The total cosmic-ray energy in the middle aged SNRs {such as} W44 and W28 is estimated to be 10$^{49}$ erg in the hadronic scenario \citep[e.g.,][]{2013ApJ...768..179Y,2017AIPC.1792d0039Y}. In addition, it is likely that escaping cosmic rays can be {more significant than accelerated cosmic-rays inside the SNR}. {In} W44, \cite{2012ApJ...749L..35U} showed the nearby clouds outside the SNR {are} $\gamma$-ray bright, and may involve a significant amount of cosmic rays {on} the order of (0.3--3)$ \times 10^{50}$ erg. {This value corresponds to $\sim$10\% of the typical kinematic energy released by a supernova explosion ($\sim$10$^{51}$ erg)}. Therefore, SNRs {such as} RXJ1713 {may} substantially contribute to the Galactic {cosmic-ray} energy budget {because} on average 10\% of the total explosion energy is needed to sustain the Galactic {cosmic-ray} flux (see \citeauthor{2013ASSP...34..221G} \citeyear{2013ASSP...34..221G} and references therein).

\subsubsection{Hadronic vs. leptonic scenarios}\label{sss:scenarios}
{The} major process {responsible for} $\gamma$-ray production in RXJ1713 {is still under debate}. The similar shell-like distributions {in} both X-rays and $\gamma$-rays make two scenarios possible. In the leptonic scenario, cosmic-ray electrons are responsible {for} both $\gamma$-ray and X-ray production. In the hadronic scenario, the shell-like ISM distribution produces shell-like $\gamma$-rays and the shock--cloud interaction produces a shell-like distribution {owing to} the enhanced nonthermal X-rays around the dense cores. A possible difference between the two is that the shock--cloud interaction causes spatial offsets {on} the order of 0.1--0.4 pc between the cores and the $\gamma$-ray peaks, which are {still} below the {current} limit of $\gamma$-ray resolution.

\begin{figure*}[t]
\begin{center}
\includegraphics[width=165mm,clip]{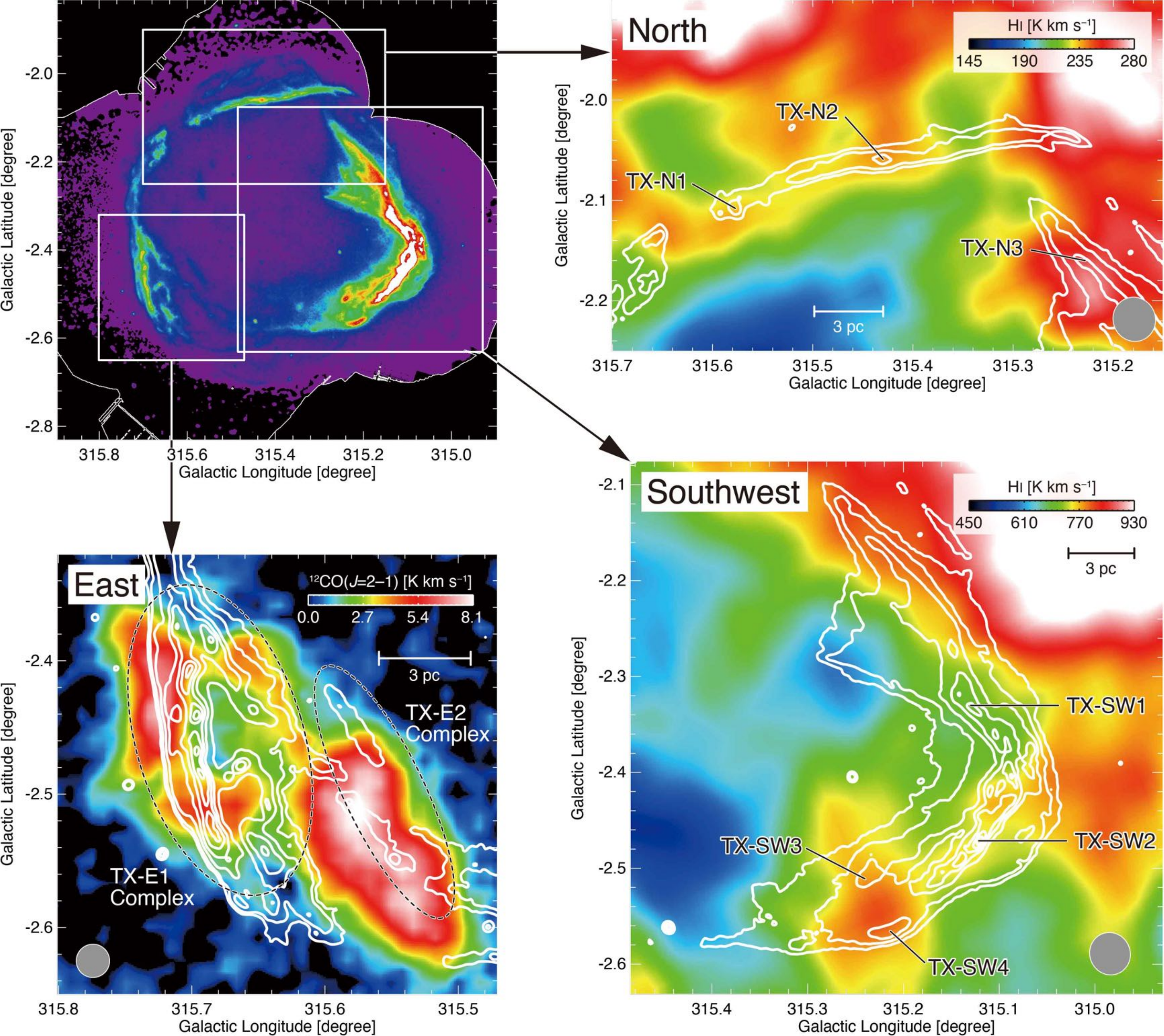}
\end{center}
\caption{{\it Top left}: Thermal X-ray image of RCW~86 in the energy band 0.5--1.0 keV. {\it Other panels}: Maps of H{\sc i} ({\it right panels}, North and Southwest) and $^{12}$CO($J$ = 2--1) ({\it {bottom-left} panel}, East) obtained with ATCA $\&$ Parkes and NANTEN2 (color scale) superposed {on} the X-ray contours. From \cite{2017JHEAp..15....1S} with permission}
\label{fig:11}
\end{figure*}

{B}road-band fitting of the $\gamma$-ray spectrum {can be} reconciled with the both scenarios according to {a} recent analysis {by} \cite{2018A&A...612A...6H}. {They} argued that both the leptonic and hadronic scenarios can explain the $\gamma$-ray observations {obtained} by H.E.S.S., based on a spectral analysis of the $\gamma$-ray observations from 2003 to 2018. {In this} work, the H.E.S.S. $\gamma$-rays {and} {\it{Suzaku}} X-rays {were combined} and {the} broad-band energy spectrum from 1 keV to 10 TeV {was presented}. The spatial resolution was conservatively set to 3.2 pc, which was adopted {for} the {\it{Suzaku}} analysis \citep{2008ApJ...685..988T}, while the new H.E.S.S. data achieved a {nominal} resolution of 0.6 pc. In the leptonic scenario, the cosmic-ray electrons are responsible for the $\gamma$-ray production. {The e}lectrons are cooled by synchrotron energy loss{, so the} magnetic field has to be small ($\sim$10 $\mu$G) for the leptonic scenario to work. {In} shock--cloud interactions, the field may become as strong as 0.1--1 mG locally, {although it may not be so strong} over a large volume. \cite{2018A&A...612A...6H} did not make a comparison with the ISM{, which} remains to be done in {the} future.

\section{RCW~86}\label{s:rcw86}
\subsection{X-rays and the ISM}\label{ss:xrays}
\begin{figure}[h]
\begin{center}
\includegraphics[width=73mm,clip]{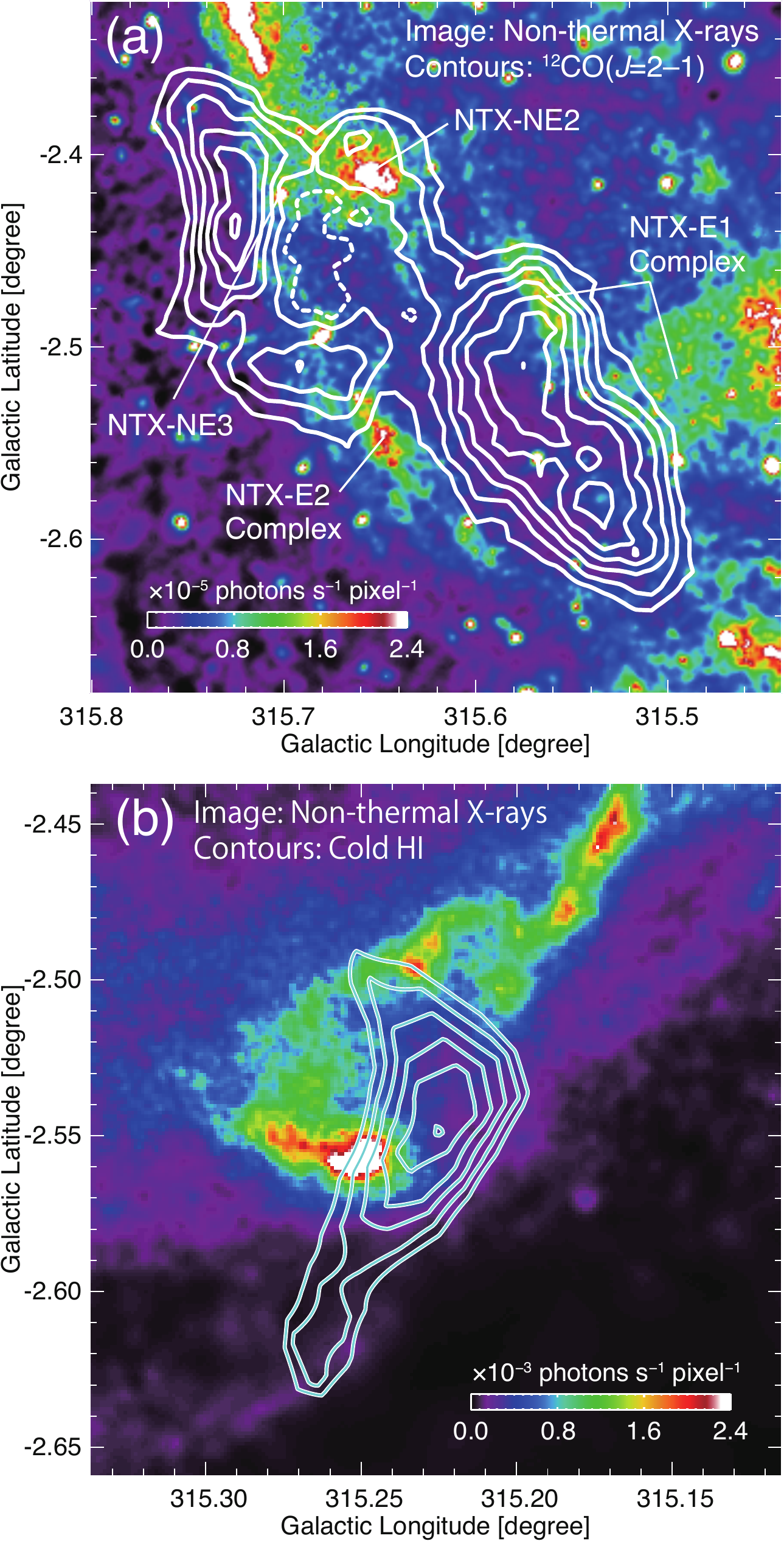}
\end{center}
\caption{{Nonthermal X-ray images ($E$: 2.0--4.5 keV) toward {the} (a) East and (b) Southwest. Superposed contours indicate (a) CO and (b) cold H{\sc i}.} From \cite{2017JHEAp..15....1S} with permission}
\label{fig:13}
\end{figure}

{The} Type Ia SNR RCW~86 is located at ($l$, $b$) = (315.4, $-2.3$) and {it exhibits} both thermal and nonthermal X-rays (Figure \ref{fig:10}a). The distance {has been} determined to be $\sim$2.5 kpc {using} several methods \citep[e.g.,][]{1969AJ.....74..879W,1996A&A...315..243R,2013MNRAS.435..910H}, and its $\sim$1800 yr {age is} based on {a} possible counterpart to the historical supernova event SN~185 recorded {in} Chinese literature in 185~AD \citep{2002ISAA....5.....S,1997AJ....114.2664S,2001ApJ...546..447D,2006ApJ...648L..33V}. The ISM in {the} velocity range from $-46$ to $-28$ km s$^{-1}$ is associated with the SNR, as found {from} the spatial correlation with the X-rays \citep{2016ApJ...819...98A,2017JHEAp..15....1S}. RCW~86 is close to {a} CO cloud, {that} is part of {the} supershell GS 314.8$-$0.1$-$34 identified in CO emission by \cite{2001PASJ...53.1003M}. \cite{2017JHEAp..15....1S} compared the X-rays with CO and H{\sc i} for two energy bands, 0.5--1.0 and 2.0--4.5 keV, where the low energy band is dominated by thermal X-rays and the high energy band by nonthermal X-rays \citep[e.g.,][see also Figure \ref{fig:10}a]{2002ApJ...581.1116R,2016ApJ...819...98A}. 

The ISM surrounding the SNR is dominated by H{\sc i}. The overall distribution of {both} H{\sc i} and CO is shell like, and the X-rays are distributed within the cavity (Figure \ref{fig:10}b). In addition, extended weak H{\sc i} emission is found inside the shell. According to \cite{2017JHEAp..15....1S}, the thermal X-ray filaments with a typical thickness of 1 pc show tight correlations with H{\sc i} and CO (Figure \ref{fig:11}), {while} the nonthermal X-rays are associated with low-density regions of H{\sc i}{, with densities} around $4 \times 10^{21}$ cm$^{-2}$. The ISM density in RCW~86 is lower than {that} in RXJ1713, which is consistent with a Type Ia SNR. Only a CO cloud with a size of 5 pc $\times$ 3 pc interacts with the shock, as verified by the high temperature estimated {from the} high ratio of the two CO transitions, CO $J$ = 2--1 and 1--0. {N}o heating sources such as {\it{IRAS / AKARI}} infrared point sources or high-mass stars {exsit} in these regions, except for the shock front of RCW~86 \citep{2017JHEAp..15....1S}.

\begin{figure*}[t]
\begin{center}
\includegraphics[width=\linewidth,clip]{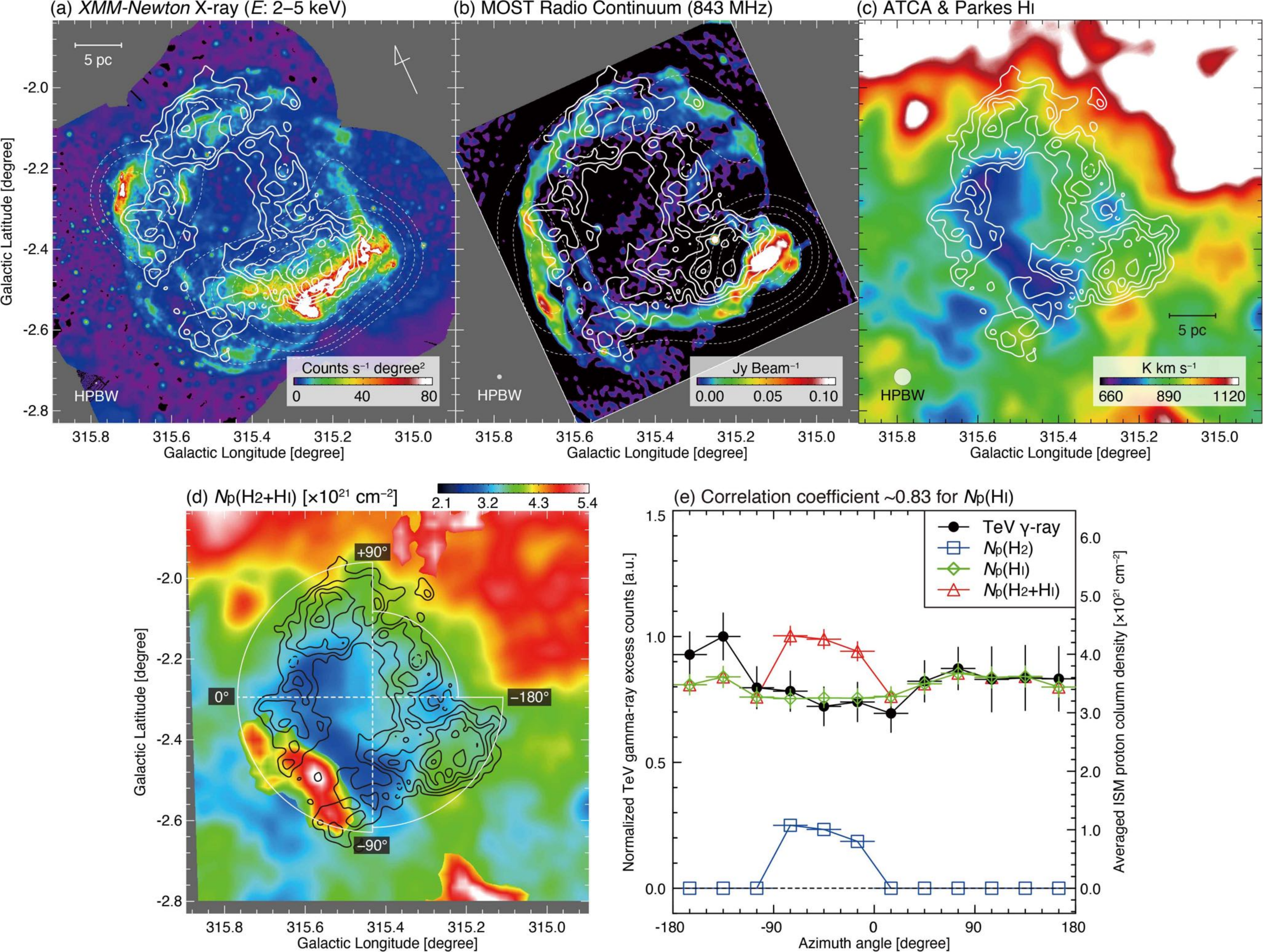}
\end{center}
\caption{(a--b) Synchrotron radiation images {from} (a) {\it XMM-Newton} X-rays ($E$: 2--5 keV) and (b) {Molonglo Observatory Synthesis Telescope (MOST)} 843 MHz radio continuum{, both} toward RCW~86. {The s}uperposed solid contours represent the H.E.S.S. TeV $\gamma$-rays. The dashed contours indicate {the} intensity distributions for each map {smoothed} with the same {point-spread function} as the $\gamma$-ray image. (c) H{\sc i} map of RCW~86 obtained with ATCA $\&$ Parkes superposed on the TeV $\gamma$-ray contours. (d) Distribution of $N_\mathrm{p}$(H$_2$ + H{\sc i}) toward RCW~86 {superposed on the TeV $\gamma$-ray contours.} (e) Azimuthal distributions of $\gamma$-rays, $N_\mathrm{p}$(H$_2$), $N_\mathrm{p}$(H{\sc i}), and $N_\mathrm{p}$(H$_2+$H{\sc i}), which are averaged {over} every 30$\arcdeg$ in {azimuthal} angle within the region enclosed by {the} white {circle} in Figure \ref{fig:14}(d). From \cite{2019ApJ...876...37S}, reproduced by permission of the AAS}
\label{fig:14}
\end{figure*}

{Because} the thermal X-rays are produced by shock heating of neutral gas {with a} density less than 100 cm$^{-3}$, the correlation with the H{\sc i} gas is a natural outcome of the shock heating. The critical difference between RCW~86 and RXJ1713 is probably the total mass of H{\sc i} gas inside the SNR shell. For RXJ1713, dense molecular clouds with {densities greater} than $10^4$ cm$^{-3}$ {remain} without being swept up by the SNR shocks. The diffuse and intercloud H{\sc i} gas {were} completely evacuated by strong stellar winds, and then creating the very low-density cavity and the swept-up dense wall \citep[e.g.,][]{2003PASJ...55L..61F,2005ApJ...631..947M,2010ApJ...724...59S,2013PASA...30...55M}. Therefore, we {did} not find any H{\sc i} envelopes around the dense molecular clouds {in RXJ1713}, and strong thermal X-rays were not detected at the peripheries (see also Section \ref{ss:gspectra}). {On the other hand,} RCW~86 shows an H{\sc i} envelope surrounding the molecular cloud in the east, and a large amount of diffuse H{\sc i} gas remains inside the SNR (see Figure \ref{fig:10}b). {This} difference is explained by {the} hypothesis that accretion winds (or disk winds) from the progenitor system of RCW~86 {are} weaker than {the} stellar winds from the high-mass progenitor of RXJ1713.

\cite{2017JHEAp..15....1S} also studied {the} association of the ISM with nonthermal X-rays{. Figure \ref{fig:13} shows nonthermal X-ray images superposed on the CO and cold H{\sc i} contours toward the east and southwest regions. The CO and H{\sc i} clumps are limb-brightened in nonthermal X-rays, which are dominated by synchrotron radiation. This indicates that shock--cloud interactions {occur} in the surface {layers} of dense clumps, where the magnetic filed is strongly amplified.} RCW~86 has no clumpy CO inside the shell, and {the gas density} seems to be smoother than in RXJ1713. The authors interpreted the {lower} ISM mass {to be} related to the older evolutionary state of the region after star formation because RCW~86 is part of a molecular supershell with an age of {$\sim$2~Myr} \citep{2001PASJ...53.1003M}. {By} contrast, in RXJ1713 we see ongoing star formation within the SNR shell, as evidenced by the protostellar outflow in CO peak ``C'' and a few more YSO candidates \citep{2010ApJ...724...59S}.

\subsection{The origin of the $\gamma$-rays}\label{ss:rcw86gamma}
Figures \ref{fig:14}(a) and \ref{fig:14}(b) show overlays of the synchrotron X-ray and radio continuum distributions on the TeV $\gamma$-ray contours{, where the lowest {$\gamma$-ray} contour level of $\gamma$-ray corresponds to a significance level of $\sim$$5\sigma$ \citep[see also Figure 1 of][]{2018A&A...612A...6H}}. {In these} images, the synchrotron emission appears as a nearly circular shell, whereas the $\gamma$-ray emission seems to deviate from this shape, particularly toward the West, where the $\gamma$-ray emission {appears to be} shifted toward the SNR interior. If true, this would {argue} against the leptonic scenario. However, {by} employing a quantitative analysis {that uses} radial profiles in the different regions of interest, \cite{2018A&A...612A...6H} showed that within statistical errors, {there is no significant difference} between {the} TeV $\gamma$-rays and {the} X-ray synchrotron emission. Future $\gamma$-ray observations {with} high sensitivity and high angular resolution are needed to confirm the spatial match or {the difference between} the two images.

\begin{figure*}[t]
\begin{center}
\includegraphics[width=\linewidth,clip]{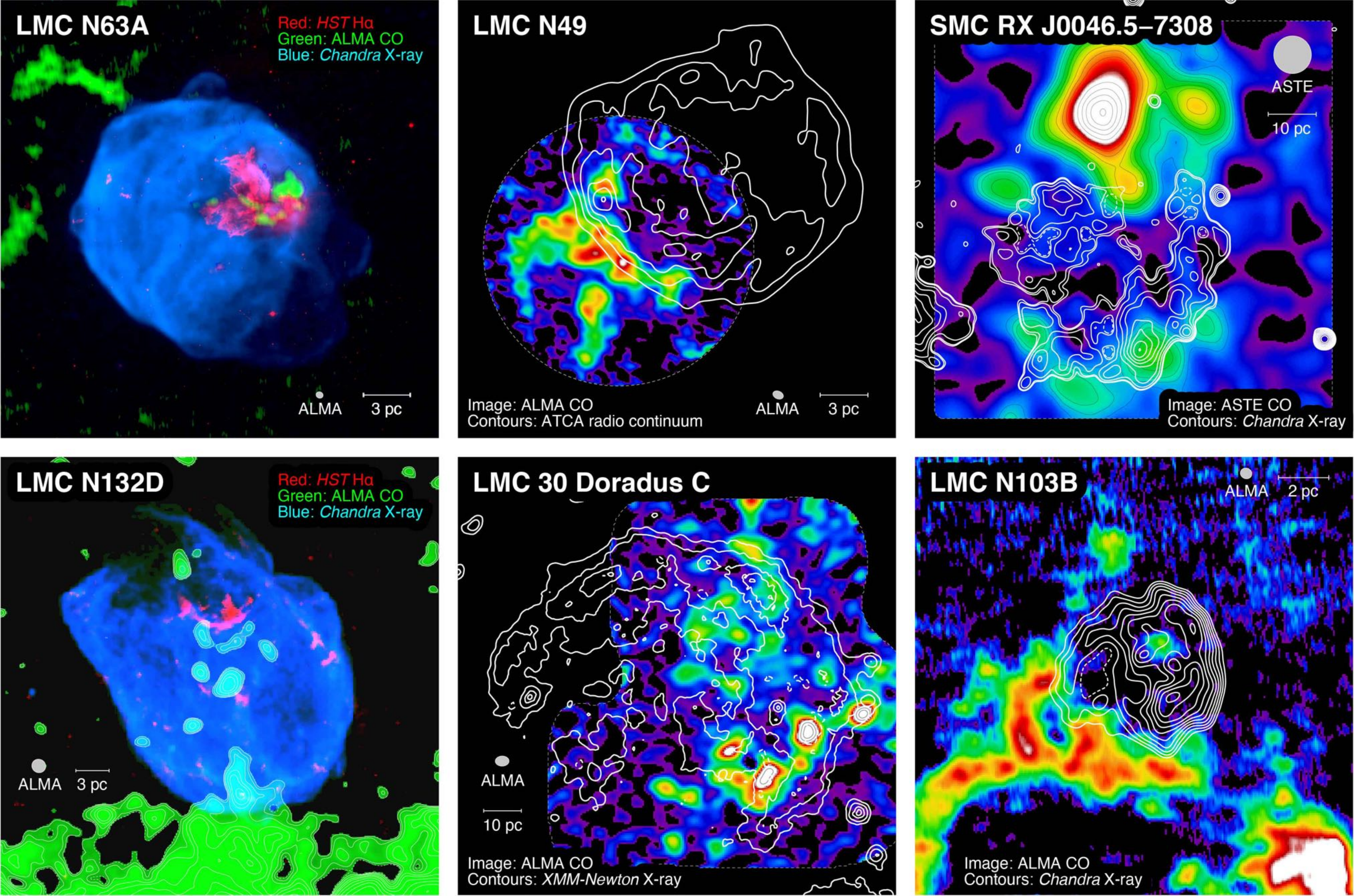}
\end{center}
\caption{CO results toward the five Magellanic SNRs N63A \citep{2019ApJ...873...40S}, N49 \citep{2018ApJ...863...55Y}, RX~J0046.5$-$7308 \citep{2019ApJ...881...85S}, N132D {\citep{2020ApJ...902...53S}}, N103B \citep{2018ApJ...867....7S}, and the Magellanic superbubble 30~Doradus~C \citep{2021arXiv210609916Y}. The CO observations of {the} LMC SNRs N63A, N49, N132D, N103B, and the LMC superbubble 30~Doradus~C were {obtained} using ALMA, while the SMC SNR RX~J0046.5$-$7308 was observed using the Atacama Submillimeter Telescope Experiment (ASTE). From \cite{2018ApJ...867....7S,2019ApJ...873...40S,2019ApJ...881...85S,2020ApJ...902...53S} and \cite{2018ApJ...863...55Y,2021arXiv210609916Y}, reproduced by permission from the AAS}
\label{fig:15}
\end{figure*}

A spatial comparison between the ISM and $\gamma$-rays therefore {remains} essential {for understanding} the origin of the $\gamma$-rays. In Figure \ref{fig:14}(c) the TeV $\gamma$-ray distribution is overlaid on the H{\sc i}, {and the comparison} shows good spatial correspondence. Figure \ref{fig:14}e shows a plot of the azimuthal distributions {of} the $\gamma$-rays and {of} the total proton column density $N_\mathrm{p}$(H$_2 + $H{\sc i}), derived using both the H{\sc i} and {the} CO. The $\gamma$-rays show good correspondence {with} the H{\sc i} {distribution}, while no enhancement {is seen} toward the CO cloud (Figures {\ref{fig:14}d and \ref{fig:14}e} in the azimuth angles from $-90^{\circ}$ to $0^{\circ}$). The poor correlation with the CO is ascribed to the effect of cosmic-ray exclusion {due to} the small penetration depths \citep[less than 1 pc;][see Section {\ref{ss:gspectra}}]{2019ApJ...876...37S}. {More precisely, the {radii of the} shock-interacting molecular clouds {constitute} a key difference between RXJ1713 and RCW~86. For RXJ1713, the cloud radius is $\sim$1 pc or less; hence, the molecular clouds play a significant role as targets of cosmic-ray protons. {However}, the cloud radius of RCW~86 is more than 5 pc \citep[][]{2017JHEAp..15....1S}. Additionally, {the} weak disk winds from the progenitor system of RCW~86 {were} not {able to} strip off {the} H{\sc i} envelope {from} the {surface of} molecular cloud. {Therefore, c}osmic-ray protons accelerated in RCW~86 can interact only with the low-density H{\sc i} gas on the surface of the molecular cloud.}

For the {azimuthal angles} from $-180^{\circ}$ to $-120^{\circ}$, we find $\gamma$-ray excesses relative to the total proton column density. We see a hint of the $\gamma$-ray excess by a factor of 1.2, which may be ascribed to the contribution {from a} leptonic $\gamma$-ray component, because bright synchrotron radiation is detected toward the SE region. The possible contribution of leptonic $\gamma$-rays to the total $\gamma$-rays {{may} be as low as 5\% if the arguments {given above} are correct.} The total cosmic-ray energy in RCW~86 is estimated to be $\sim$$1 \times 10^{48}$ erg. A similar argument to the case of RXJ1713 is perhaps applicable here, and the energy is a typical one for a young SNR with an age of $\sim$2000 yr. Considering the lower sensitivity of the current H.E.S.S. data {for} RCW~86, we do not discuss the $\gamma$-ray origin {further} until more sensitive $\gamma$-ray data become available. 

\section{Summary and Future Prospects}\label{s:summary}
We {have} reviewed the ISM associated with young supernova remnants and its connection with the {observed} X-rays and $\gamma$-rays. {These} high-energy radiations are closely related to cosmic rays {that} are probably accelerated via DSA. The present focus was on the young, bright TeV $\gamma$-ray SNRs RXJ1713 and RCW~86. Shock--cloud interactions in {a} clumpy ISM play an important role in regulating the X-ray properties as well as the $\gamma$-ray production.

The pursuit of the ISM in SNRs is being extended to nearby galaxies, including the Large Magellanic Cloud (LMC) and the Small Magellanic Cloud (SMC). The ISM associated with the SNRs {has been} identified in many of them, providing promising candidates for further studies on the role of the ISM \citep[][]{2015ASPC..499..257S,2017ApJ...843...61S,2017AIPC.1792d0038S,2018ApJ...867....7S,2019ApJ...873...40S,2019ApJ...881...85S,2020ApJ...902...53S,2018ApJ...863...55Y,2021ApJ...under.review,2018MNRAS.479.1800R,2019MNRAS.486.2507A}. The smaller contamination along the line of sight is advantageous for identifying molecular clouds associated with the SNRs and for using optical and infrared datasets without strong stellar absorption. Although the Magellanic SNRs {lie at} 50--60 times larger distances from us than {compared with the distance of} RXJ1713 {from us} \citep[e.g.,][]{2005MNRAS.357..304H,2013Natur.495...76P}, ALMA's unprecedented sensitivity and spatial resolution {have} enabled us to resolve spatially {the} cloud-scale structures of the ISM associated with the Magellanic SNRs.

Figure \ref{fig:15} shows CO mapping results toward the five Magellanic SNRs and a superbubble that are bright in X-rays. The molecular clouds {lie} nicely along, or embedded within, the SNR shells, indicating {that} the shock--cloud interactions {have} occurred. 30~Doradus~C, one of the brightest X- and $\gamma$-ray superbubble{s} in the Local group {(see \citeauthor{2020Ap&SS.365....6K} \citeyear{2020Ap&SS.365....6K} and references therein)}, provides {one of} the best {laboratories} for studying shock--cloud interactions because of its large apparent diameter and very bright TeV $\gamma$-rays \citep[][]{2015Sci...347..406H}. Future observations using ALMA, {\it{Chandra}}, {and the} CTA will allow us to study in detail not only shock--cloud interactions but also the hadronic $\gamma$-ray radiation from the Magellanic SNRs.

We summarize the main points of {this} review as follows:

\begin{enumerate}
\item {Spatial comparisons} between X-rays and the ISM provide a powerful {approach for determining the distances} to {SNRs,} which {is otherwise difficult for} SNRs in the Galactic plane {because of} heavy {obscuration}. {For RXJ1713, the ISM delineates the outer boundary of the nonthermal X-ray shell at pc scales, while X-ray bright spots (or filaments) are anti-correlated} with the ISM {clouds/cloudlets at small-scales (sub-pc scales)}, allowing a robust distance estimate. RCW~86 is an easier case, for which distance {has been} determined accurately {owing} to {its} relatively high Galactic latitude.

\item When we consider the ISM surrounding {the} SNRs, it is important to {recognize} that the neutral ISM is characterized by highly clumped {distributions, with} clumps {that} are much denser than the nominal uniform density {of} 1 cm$^{-3}$ assumed in many previous works. The ISM shell in a SNR {surrounds} a low-density cavity, and the shell {comprises} a layer {that} includes {tiny,} dense clumps with a small volume filling factor. {Consequently, an} SNR has a large low-density volume, where DSA works, as well as dense clumps, which interact with the cosmic rays and shocks. Magnetohydrodynamical numerical simulations of shock--cloud interactions reveal that the dense clumps cause strong deformations of the shock fronts. The deformations produce highly turbulent velocity distributions, which entangle and {amplify} the magnetic field to mG {levels} \citep{2012ApJ...744...71I}. In most {parts} of the low-density cavity, however, the shocks propagate with little deceleration or deformation.

\item The clumpy ISM picture requires {modification of} the previous {scenarios that were} based on {a} homogeneous ISM. The observed $\gamma$-ray spectrum of RXJ1713 is hard, according to {the} {{\it{Fermi}}} collaboration, and the spectrum {has previous been} interpreted in terms of the leptonic scenario {in a} uniform ISM picture. {However, we have} shown that {a} the clumpy ISM {can explain} the hard spectrum {{in} the GeV band ($\Gamma_\mathrm{GeV}$ = 1.5)} equally well because the penetration of cosmic rays into dense clumps is {inhibited} by diffusive scattering due to the turbulent magnetic field. We {have discussed} the implications of {a clumpy ISM for interpreting} the observations of several SNRs and {have outlined} the future prospects {for this field}.
\end{enumerate}

%
\acknowledgments
We truly appreciate the anonymous referee for the useful comments and suggestions, which helped authors to improve the review. We would like to thank the editors of this Topical Collection Ralf-J\"{u}rgen Dettmar, Manami Sasaki, and Julia Tjus for inviting us to write this review. We are also grateful to Rei Enokiya for {the} useful comments, which helped the authors improve the review paper. This review paper makes use of the following ALMA data: ADS/JAO.ALMA\#2013.1.01042.S, ADS/JAO.ALMA\#2015.1.01130.S, ADS/JAO.ALMA
\#2015.1.01195.S, ADS/JAO.ALMA\#2015.1.01232.S, ADS/JAO.ALMA\#2017.1.01363.S, and ADS/JAO.AL
MA\#2017.1.01406.S. ALMA is a partnership of ESO (representing its member states), NSF (USA) and NINS (Japan), together with NRC (Canada), MOST and ASIAA (Taiwan), and KASI (Republic of Korea), in cooperation with the Republic of Chile. The Joint ALMA Observatory is operated by ESO, AUI/NRAO and NAOJ. This review was supported by JSPS KAKENHI Grant Numbers JP15H05694 (YF), JP19K14758 (HS), JP19H05075 (HS), {21H01136 (HS)}. {The authors would like to thank Enago (www.enago.jp) for the English language review.}



%

%

\end{document}